\documentclass[aps,twocolumn,showpacs,amsmath,amssymb,superscriptaddress]{revtex4-1}
\usepackage[utf8]{inputenc}

\usepackage{tikz}
\usepackage{graphicx}
\usepackage{amsfonts}
\usepackage{amsmath}
\usepackage{amssymb,bbold}
\usepackage{color}
\usepackage{bm}
\usepackage{bbm}
\usepackage{enumerate}
\usepackage{amsfonts, amsmath, amsthm, amssymb} 
\usepackage{mathtools}
\usepackage{array}
\usepackage[colorlinks,bookmarks=false,citecolor=blue,linkcolor=red,urlcolor=blue]{hyperref}

\begin{document}

\title{Quench dynamics of quantum spin models with flat bands of excitations}

\author{Rapha\"el Menu}
\affiliation{Univ Lyon, Ens de Lyon, Univ Claude Bernard, CNRS, Laboratoire de Physique, F-69342 Lyon, France}
\author{Tommaso Roscilde}
\affiliation{Univ Lyon, Ens de Lyon, Univ Claude Bernard, CNRS, Laboratoire de Physique, F-69342 Lyon, France}
\affiliation{Institut Universitaire de France, 103 boulevard Saint-Michel, 75005 Paris, France}

\date{\today}

\begin{abstract}
We investigate the unitary evolution following a quantum quench in quantum spin models possessing a (nearly) flat band in the linear excitation spectrum. Inspired by the perspective offered by ensembles of individually trapped Rydberg atoms, we focus on the paradigmatic trasverse-field Ising model on two dimensional lattices featuring a flat band as a result of destructive interference effects (Lieb and Kagom\'e lattice); or a nearly flat band due to a strong energy mismatch among sublattices (triangular lattice). Making use of linear spin-wave theory, we show that quantum quenches, equipped with single-spin imaging, can directly reveal the spatially localized nature of the dispersionless excitations, and their slow propagation or lack of propagation altogether. Moreover we show that Fourier analysis applied to the post-quench time evolution of wavevector-dependent quantities allows for the spectroscopic reconstruction of the flat bands. Our results pave the way for future experiments with Rydberg quantum simulators, which can extend our linear spin-wave study to the fully nonlinear regime, characterized by the appearance of dense, strongly interacting gases of dispersionless excitations. 
\end{abstract}

\maketitle

\section{Introduction}

 \emph{Quantum quenches in many-body systems.} The non-equilibrium unitary dynamics of closed quantum many-body systems represents one of the most active topics of research in modern condensed matter \cite{Polkovnikovetal2011,d'Alessioetal2016}, largely inspired by the impressive experimental progress in the coherent manipulation of model systems (or quantum simulators \cite{Georgescuetal2014}) such as strongly interacting trapped atoms \cite{GrossB2017,Altman2015,Monroeetal2014}, superconducting-qubit architectures \cite{Roushanetal2017}, etc. A closed quantum many-body system, prepared in a state which is not an eigenstate of its evolution Hamiltonian ${\cal H}$, undergoes a so-called quantum quench, which triggers a subsequent process of relaxation, namely the complex reorganization of the entanglement and correlation patterns in the many-body wavefunction \cite{Eisertetal2015}. 
 
 A central aspect determining the quantum-quench dynamics is given by the nature of the excited states of ${\cal H}$ -- and quantum quenches may invoke excited states which are arbitrarily high in the spectrum. Typically our knowledge (both theoretical and experimental) about the spectrum of many-body systems is limited to elementary excitations, generically described as free quasiparticles; and it focuses on the dispersion relation, namely on the momentum-frequency ``portrait" of excitations, reconstructed spectroscopically within the linear-response regime. In this respect, quantum quenches in closed quantum systems provide a new paradigm, because they probe the dynamics of excitations in \emph{real} time, and, when access to the microscopic degrees of freedom is available, also in \emph{real} space. This represent a unique opportunity to understand the dynamical consequences of different types of excitations (dispersive \emph{vs.} dispersionless, extended \emph{vs.} localized, etc.), and in particular their role in the spatial spreading of entanglement and correlations which is necessarily implied by a quench \cite{Eisertetal2015}. 

 In particular, \emph{local} quenches -- namely evolutions of initial states which are only locally perturbed with respect to an eigenstate of ${\cal H}$ -- probe the real-time/real-space spreading of initially localized wavepackets of excitations. On the other hand, \emph{global} quenches -- namely non-equilibrium evolutions of homogeneous initial states -- probe the dispersion relation of excitations via the mechanism by which two-point correlations rearrange in the system. Indeed the rearrangement of correlations is expected to occur within a causal light cone, whose aperture and internal features are dictated by the group velocities of the elementary excitations \cite{calabrese_evolution_2005, Cevolanietal2017}. Several recent studies, both experimental \cite{Cheneauetal2012, Langen2013} and theoretical \cite{Barmettleretal2012,hauke_spread_2013,Carleoetal2014,cevolanietal2016}  have established the quantitative relationship between the properties of elementary excitations and the space/time features of the quench dynamics. 

\begin{center}
\begin{figure}[t]
\includegraphics[width=0.8\columnwidth]{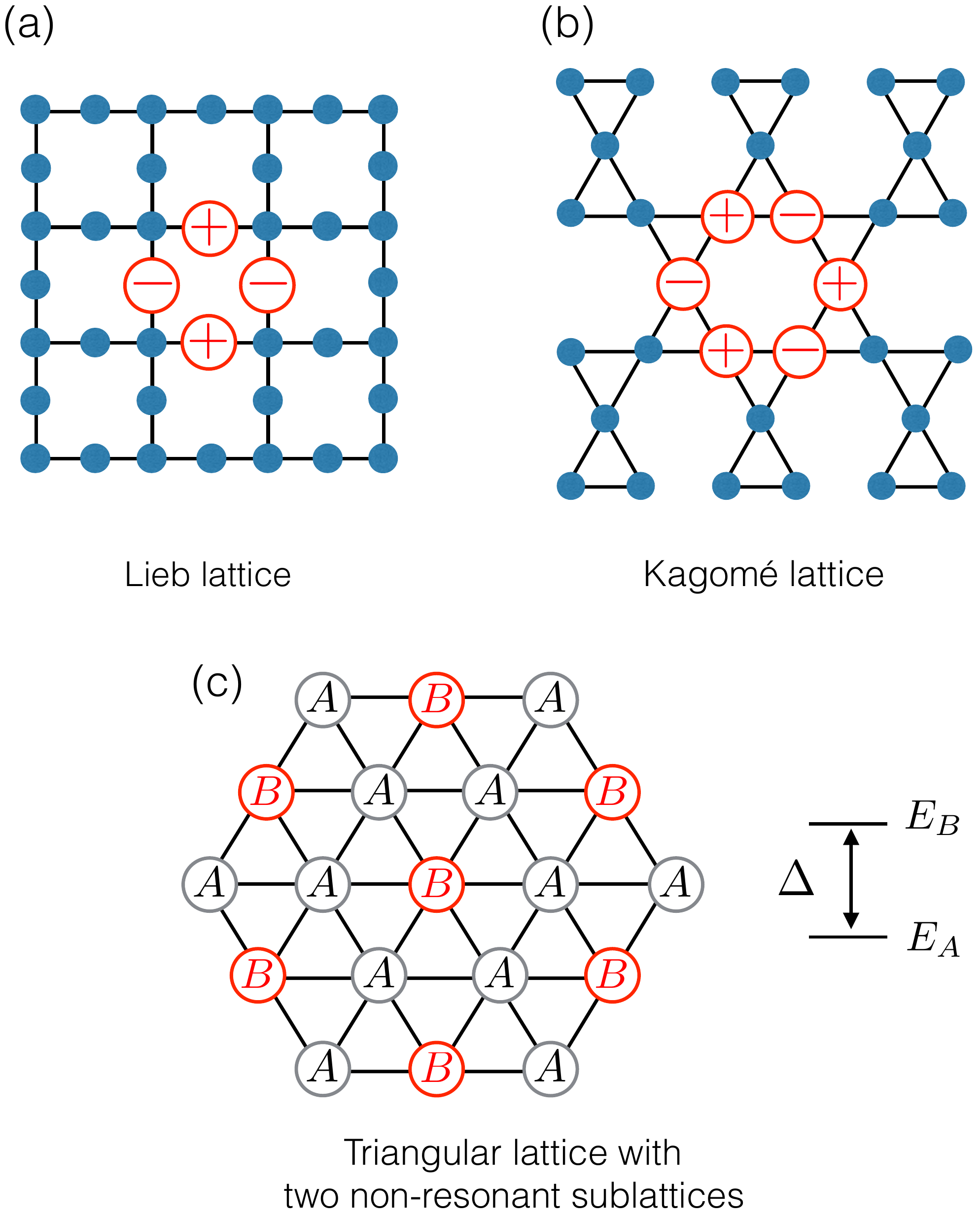}
\caption{Lattice geometries featuring (nearly) flat bands in the case of tight-binding models. (a)-(b) The Lieb and Kagom\'e lattice feature localized single-particle modes (indicated in red in the figure) with amplitudes of uniform magnitude but alternating sign, defining $\oplus$-sites and $\ominus$-sites. The propagation of these modes displays fully destructive interference in all directions (or ''Aharonov-Bohm caging"), because each site neighboring the support of the localized mode is connected to as many $\oplus$-sites as $\ominus$-sites. These $N/3$ localized modes, albeit non-orthogonal, span one of the three bands of the lattices in question, which is therefore completely flat. (c) If the $A$ and $B$ sublattices of a triangular lattice, indicated in the figure, are offset by an energy $\Delta \gg J$, $J$ being the hopping amplitude, the spectrum is composed of two bands of width $\sim J$, and a band of width $\sim J^2/\Delta \ll J$. The modes of the latter band overlap predominantly with the $B$ sublattice, $J^2/\Delta$ being the effective hopping between two $B$ sites. }
\label{flatband-sketch}
\end{figure}
\end{center}

 \emph{Flat-band systems.} 
 Most of the recent experimental and theoretical studies have focused on lattice models with a Bravais lattice, featuring a single band of elementary quasiparticle excitations. Such models generically admit a well-defined maximal group velocity for the quasiparticles, controlling the light-cone aperture. Notable exceptions in this context are offered by models with long-range couplings \cite{hauke_spread_2013,cevolanietal2016,Cevolanietal2017,Frerotetal2018}, which may exhibit divergent group velocities and super-ballistic propagation of correlation fronts, as well as a complex multi-speed dynamics. 
 
In this work we extend the study of quantum quenches to the case of elementary excitations exhibiting a \emph{multi-band} structure; and, most importantly, featuring one (nearly) flat band. Flat bands are a very active topic of research within the broader subject of wave propagation in complex periodic media (see Ref.~\cite{Leykametal2018} for a recent review), ranging from superconducting circuits for Cooper pairs \cite{Teseietal2006}, to engineered atomic lattices for electrons \cite{Drostetal2017,Slotetal2017} to photonic \cite{vicencio_observation_2015,mukherjee_observation_2015} and polaritonic lattices \cite{baboux_bosonic_2016}, to optical lattices for cold atoms \cite{Taieetal2015}, among others.  At the theory level, flat bands of excitations emerge naturally in frustrated magnets exposed to strong magnetic fields \cite{Richter2005, Derzhkoetal2007} and they are also natural hosts of fascinating strong correlation phenomena such as interaction-induced ferromagnetism \cite{Tasaki1998}. If band flatness is in general a fine-tuned properties, a less strict requirement is that the bandwidth of the (nearly) flat band be much smaller than that of the dispersive bands -- a property which is robust to perturbations with a strength far smaller than the bandwidth of the other bands. In the following, unless further specifications are added, we shall generically refer to systems featuring a perfectly or nearly flat band as flat-band systems. 
 
 Flat bands can emerge due to different mechanisms. Perfect flatness can appear when the lattice structure admits so-called ``Aharonov-Bohm cages" \cite{Vidaletal1998}, namely the existence of localized modes that cannot propagate because of perfect destructive interference effects. This is a well-known feature of famous lattice structures such as the Lieb, Kagom\'e and dice lattice, among others -- see Fig. \ref{flatband-sketch}(a,b). Imperfect flatness, on the other hand, can emerge if an energy offset is imposed between two inequivalent sublattices ($A$ and $B$) in which the lattice is divided, where the $A$ sites form a connected network hosting the highly dispersive modes, while the $B$ sites form a non-connected sublattice hosting the weakly dispersive modes -- see Fig. \ref{flatband-sketch}(c) for the triangular lattice. In the following we shall explore examples from both categories.

  \emph{Quantum magnetism and Rydberg quantum simulators.} Most flat-band systems of current experimental interest host either linear modes (\emph{e.g.} in photon waveguide lattices) or weakly coupled modes (\emph{e.g.} in polaritonic lattices). Here we take a different route, focusing on strongly correlated systems, namely lattice spin Hamiltonians. Such systems admit linear spin waves with flat-band dispersions as emergent elementary excitations in the low-energy regime; but at the same time arbitrary non-linearities can be triggered by increasing the energy of the initial quench, and hence the populations of the excitation modes. Our focus shall be particularly on transverse-field Ising models (TFIMs), which are not only paradigmatic models of quantum magnetism \cite{TFIMbook}, but they also faithfully describe the dynamics of individually trapped neutral atoms in periodic arrays, in which sizeable intersite interactions are triggered by exciting the atoms towards Rydberg states \cite{Browaeysetal2016}. This experimental platform represents a unique opportunity for the quantum simulation of quantum magnetism within atomic physics, as already demonstrated by recent ground-breaking experiments \cite{Labuhnetal2016, barredo_atom-by-atom_2016, lienhard_observing_2017,guardado-sanchez_probing_2017,Bernienetal2017}. The remarkable flexibility in the geometry of the array allows to realize arbitrary two- an three-dimensional lattices \cite{barredo_atom-by-atom_2016, barredo_synthetic_2017}, including the ones that are relevant to the present work. Finally, single-spin addressability offered by Rydberg quantum simulators allows to trigger both global and arbitrary local quantum quenches, and to fully reconstruct the local magnetization profile as well as spin-spin correlations. 
 
 \emph{Quench dynamics of quantum Ising models with flat bands.} The present work is concerned with the study of quench dynamics in quantum Ising Hamiltonians in the regime of small quenches, which allows for  a quantitative treatment based on the linear spin-wave approach.  
 Having a band whose width is well separated in energy from that of all other bands introduces an inherent multi-speed structure, as well as a peculiar spatial structure, to the quench dynamics of correlations. Indeed we find that after global quenches, the spreading of correlations on the triangular lattice shows a fast buildup of spin-spin correlations among $A$ sites, and between $A$ and $B$ sites (see Fig.~\ref{flatband-sketch}(c)), but a very slow one between $B$ sites, as the latter is primarily driven by the flat-band modes. 
 The existence of localized modes is even more dramatically revealed by local quenches: in the particular case of the Lieb and Kagom\'e lattice, the magnetization which is injected on one site in the initial state remains Aharonov-Bohm caged, namely it is trapped by the localized states that overlap with the site in question.      
 In all the systems of interest, the Fourier analysis of the post-quench evolution of wavevector-dependent quantities allows for a spectroscopic reconstruction of the flat bands. This implies that quantum quenches on quantum simulators with single-site detection provide a unique insight into the joint wavevector/frequency as well as space/time structure of the flat-band modes.

\emph{Summary.}
 The structure of the paper is as follows. Sec.~\ref{s.Ising} illustrates the application of linear spin-wave theory to the quench dynamics of quantum Ising models; Sec.~\ref{s.triangular} discusses our results for the triangular lattice, while Sec.~\ref{s.LiebKagome} focuses on the Lieb and Kagom\'e lattice. Conclusions and an outlook are provided in Sec.~\ref{s.conclusion}.
 
\section{Quantum quenches in transverse-field Ising models}
\label{s.Ising}
\subsection{Model Hamiltonian}

 We focus our attention on $S=1/2$ TFIMs in two dimensional lattices, which can faithfully model the dynamics of arrays of individually trapped Rydberg atoms \cite{Browaeysetal2016}, as well as of other quantum simulation platforms, such as trapped ions \cite{Monroeetal2014} or superconducting circuits \cite{Johnson2011}. More specifically we will consider a lattice of $N$ sites, which can be paved with unit cells consisting of $m$ sites. Be $l$ the index of the unit cell, and $p$ the index of the site on the unit cell, so that the couple $(lp)$ uniquely identifies a site on the lattice.  
  The matrix providing the coupling among sites $i$ and $j$ can be conveniently rewritten as $J^{ll'}_{pp'}$, where $i = (lp)$ and $j = (l'p')$ -- namely the $N\times N$ coupling matrix is decomposed into $m\times m$ submatrices $J^{ll'}$ providing the couplings between the $l$-th and $l'$-th unit cell.  

 The Hamiltonian of the system reads then 
 \begin{equation}
    \label{e.Ham}
    \mathcal{H}=  \dfrac{1}{2}\sum_{lp;l'p'} J^{ll'} _{pp'} S^{z}_{lp}S^{z}_{l'p'}-\Gamma\sum_{lp}{S^{x}_{lp}} - H \sum_{lp} {S^{z}_{lp}}
\end{equation}
where $S^{\alpha}$ are spin-$S$ operators ($\alpha = x, y, z$), and we have included a transverse field $\Gamma$ and a longitudinal field $H$. We shall conduct the discussion for the general case of spin-$S$ spins, but specialize all the results shown in this work to the case $S=1/2$.

In Rydberg quantum simulators the two internal states encoding the $S=1/2$ spin can be represented by the ground state and a highly excited Rydberg state, coupled by a radiation with Rabi frequency $\Gamma$, and whose detuning from the atomic transition is expressed by $H$. Under this circumstance two atoms at distance $r$ which  have been both excited to a Rydberg state interact via a repulsive van-der-Waals interaction $J(r) = C_6/r^6$, rapidly decaying with the distance. Unless otherwise specified, in the following we will neglect all couplings beyond nearest-neighbor ones, with strength $J = C_6/d^6$ ($d$ being the lattice spacing). Including the long-range tail would only minimally alter our results, as we shall see explicitly. Due to the repulsive nature of the interactions, the natural spin-spin couplings are antiferromagnetic, namely $J>0$.

\subsection{Quenches starting from mean-field states}
\label{s.MF}

 The study of non-equilibrium dynamics in strongly interacting spin models such as Eq.~\eqref{e.Ham} poses considerable theoretical challenges, and in general terms it can only be tackled numerically via exact diagonalization on small clusters \cite{Labuhnetal2016} or via variational approaches \cite{BlassR2016}. The challenges raised by the simulation of the unitary dynamics generated by  $\exp(-i{\cal H}t)$ represent one of the strongest motivations for the experimental effort of quantum simulation based on the above mentioned platforms.
 
 In order to address the study of the unitary dynamics semi-analytically, we shall restrict our attention to situations in which the static as well as dynamic properties of  Eq.~\eqref{e.Ham} can be faithfully described within linear spin-wave (LSW) theory, which amounts to approximating the non-linear Hamiltonian by a collection of harmonic oscillators describing quantum fluctuations around the mean-field (MF) solution. This approach proves extremely successful in studying the dynamics of lattice spin systems, as demonstrated by numerous recent studies \cite{hauke_spread_2013,cevolanietal2016,Cevolanietal2017,Frerotetal2018}, and it relies on the following essential condition: the unitarily evolved state of the system must remain ``close" (in a way to be specified below) to a MF state. MF states are generically factorized states of the form $\vert\Psi_{\rm MF}\rangle=\bigotimes_{lp}{\vert\psi_{lp}\rangle}$, and they are extremely natural in the context of quantum simulation, as they provide the best fiducial states into which the system is initialized before the quench dynamics starts. For $S=1/2$ spins, the single-spin state  $\vert\psi_{lp}\rangle$ can also be indicated as 
 \begin{equation}
 |\theta_{lp},\phi_{lp}\rangle = \cos(\theta_{lp}/2) |\uparrow\rangle + 
 e^{i\phi_{lp}} \sin(\theta_{lp}/2) |\downarrow\rangle
 \end{equation}
 namely a state whose Bloch vector points along the $(\theta_{lp},\phi_{lp})$ direction on the unit sphere.

In the following, we shall consider two special instances of initial MF states: 
\begin{itemize}
\item 1) the MF approximation to the ground state of ${\cal H}$, $|\Psi_{\rm MF}^{(+)}\rangle$ ; and
\item 2) the MF approximation to the ground state of $-{\cal H}$, $|\Psi_{\rm MF}^{(-)}\rangle$, namely to the most excited state of ${\cal H}$.
\end{itemize}
 In case 1), the quench initializes the system in the low-energy sector of the model, assuming that the excitation energy of the MF ground state lies close to the actual ground-state energy, namely $\langle\Psi_{\rm MF}^{(+)} | {\cal H} | \Psi_{\rm MF}^{(+)} \rangle  - E_0 \ll N J$. Here $E_0$ is the true ground-state energy, and $J$ is the nearest-neighbor spin-spin coupling.  Under this assumption (which can be verified a posteriori) the evolved state  $\exp(-i~{\cal H}~t) | \Psi_{\rm MF}^{(+)}\rangle$ will remain close to the initial mean-field state, so that only harmonic fluctuations around it can be considered -- this is the central assumption of LSW theory away from equilibrium (see below for further discussion). 
 
  In case 2), one can make use of the time-reversal invariance of both ${\cal H}$ and of the MF ground state to its most excited state (namely of the fact that they admit a representation as a real-valued matrix and real-valued vector, respectively, on the same basis), to show that all expectation values calculated on the forward-evolved state $\exp(-i{\cal H}t) | \Psi_{\rm MF}^{(-)}\rangle$  are equivalent to those calculated on the \emph{backward}-evolved state, namely  $\exp(i~{\cal H}~t) | \Psi_{\rm MF}^{(-)}\rangle$ \cite{Frerotetal2018}. This means that, concerning the physically accessible observables, evolving the MF most excited state with ${\cal H}$ is equivalent to evolving with ${-\cal H}$, for which the state in question is the MF ground state. 
 The antiferromagnetic Hamiltonians we shall consider can exhibit \emph{frustration} when cast on the triangular or Kagom\'e lattice -- namely the impossibility of minimizing the various energy terms individually within a MF approach. But ultimately it is the choice of the initial state which dictates whether frustration is relevant at all to the dynamics: indeed choosing $|\Psi_{\rm MF}^{(-)}\rangle$ amounts to effectively evolve with a \emph{ferromagnetic} (namely unfrustrated) Hamiltonian $-{\cal H}$. This aspect will be a fundamental asset for the use of LSW theory to study the dynamics. 

\subsection{Linear spin-wave theory for the TFIM} 
\label{LSW}

 The LSW approach to the static properties of our models of interest starts with the determination of the MF ground state, namely the minimization of the variational energy 
 \begin{equation}
 E_{\mathrm{MF}}(\{\theta_{\rm lp},\phi_{\rm lp}\}) = \langle \Psi_{\rm MF} | {\cal H} | \Psi_{\rm MF} \rangle
 \end{equation}
 The MF ground state has generically a periodic structure, with a unit cell which coincides with, or exceeds, the geometric unit cell of the lattice. In the following we shall define as \emph{magnetic} unit cell the one of the MF solution, consisting of $m$ sites. The angular variables parametrizing the MF solution are then $m$ pairs of angles, 
 $\{\theta_p, \phi_p \}$ repeating themselves between unit cells. 
 
 The minimization of the MF energy defines then rotation operators ${\cal R}_p = {\cal R}(\theta_p,\phi_p)$ which rotate the $z$ axis (chosen as quantization axis) to coincide with the local spin orientation, that we shall call $z'$. In this way the rotated MF state is a polarized state along $z'$.  In the specific case of TFIMs, the Bloch vectors of the individual spins in the MF solution are all lying in the ($x$,$z$) plane (as the $y$ spin component is absent from the Hamiltonian), so that ${\cal R}$ is a rotation around the $y$ axis of an angle $\theta_p$, namely $\phi_p=0$ everywhere.  Such a rotation defines new spin operators  $\bm{S}'_{lp}=\mathcal{R}_y\left(\theta_p\right)\bm{S}_{lp}$ with 
 \begin{eqnarray}
 S'^x_{lp} &=&  \cos\theta_{lp} S^x_{lp} - \sin\theta_{lp} S^z_{lp} \nonumber \\
 S'^y &=& S^y \nonumber \\
 S'^z_{lp} & = &  \sin\theta_{lp} S^x_{lp} + \cos\theta_{lp} S^z_{lp}   
 \end{eqnarray}
The Ising Hamiltonian in terms of the new operators takes the form 
\begin{eqnarray}
{\cal H} & = & \sum_{lp;l'p'} J_{pp'}^{ll'} \cos\theta_{lp} \cos\theta_{l'p'} S'^z_{lp} S'^z_{l'p'} \nonumber \\
& + & \sum_{lp;l'p'} J_{pp'}^{ll'} \cos\theta_{lp} \sin\theta_{l'p'} S'^z_{lp} S'^x_{l'p'} \nonumber \\
& + & \sum_{lp;l'p'} J_{pp'}^{ll'} \sin\theta_{lp} \cos\theta_{l'p'} S'^x_{lp} S'^z_{l'p'} \nonumber \\
& + & \sum_{lp;l'p'} J_{pp'}^{ll'} \sin\theta_{lp} \sin\theta_{l'p'} S'^x_{lp} S'^x_{l'p'} \nonumber \\
& - & \sum_{lp} (\Gamma\cos\theta_{lp} - H \sin\theta_{lp}) S'^x_{lp}  \nonumber \\
& - & \sum_{lp} (\Gamma\sin\theta_{lp} + H \cos\theta_{lp}) S'^z_{lp} ~.
\label{e.Ham'}
\end{eqnarray}
 
 LSW theory consists then in treating harmonic fluctuations beyond the MF approximation, which are introduced by mapping the ${\bm S}'$ spins onto bosons via the Holstein-Primakoff (HP) transformation, $S_{lp}'^z = S-b_{lp}^{\dagger} b_{lp}$, $S_{lp}'^+ = b_{lp} \sqrt{2S- b_{lp}^{\dagger} b_{lp}}$. Here $b,b^{\dagger}$ are bosonic operators, $[b,b^{\dagger}]=1$, such that the boson number operator $b^{\dagger} b$ measures the deviation of the $S'^z$ spin component with respect to the perfect alignment present in the MF ground state. Linearizing the HP transformation under the assumption that $\langle b^{\dagger} b \rangle \ll 2S$ for all the states of interest -- namely taking $S'^{+}_{lp}\simeq\sqrt{2S} ~b_{lp}$, and substituting into the Hamiltonian Eq.~\eqref{e.Ham'} -- one generically obtains a bosonic Hamiltonian of the form
\begin{equation}
    \label{Eq_3}
    \mathcal{H}~=~E_{\rm MF}~+~\mathcal{H}_2~+~{\cal O}\left (\frac{b}{\sqrt{2S}} \right)^3,
\end{equation}
where $E_{\rm MF}$ is the MF ground-state energy,  and $\mathcal{H}_2$ is a quadratic form of the bosonic operators -- the linear terms in the bosonic operators disappear when expanding around the correct MF ground state.
The generic form for ${\cal H}_2$ is as follows
\begin{equation}
\label{e.H2}
\mathcal {H}_2 = \sum_{ll'} \sum_{pp'} \begin{pmatrix}  b^{\dagger}_{lp}  \\ b_{lp} \end{pmatrix}^T {\mathcal {A}}_{lp;l'p'}  \begin{pmatrix}  b_{l'p'}  \\ b^{\dagger}_{l'p'} \end{pmatrix}
\end{equation}
where $\mathcal{A}$ is a $2\times 2$ real-valued symmetrix matrix. The specific form of the MF energy and quadratic form is given in Appendix \ref{app1} for the models of interest to this work.

To diagonalize the quadratic form one then moves to Fourier space by introducing the operators $b_{\bm{k},p}=\sqrt{m/N}\sum_{l}{\exp(i\bm{k}\cdot\bm{r}_l)b_{l,p}}$; in doing so the quadratic Hamiltonian $\mathcal{H}_2$ takes a block-diagonal form with $2m\times 2m$ matrices on the diagonal
\begin{equation}
\label{e.H2'}
{\cal H}_2 = \sum_{\bm k} \begin{pmatrix} \{b^{\dagger}_{{\bm k},p} \} \\ \{ b_{-{\bm k},p} \} \end{pmatrix}^T {\cal H}_{\bm k} \begin{pmatrix} \{b_{{\bm k},p} \} \\ \{ b^{\dagger}_{-{\bm k},p} \} \end{pmatrix} 
\end{equation}
where the symbol $\{ b_{{\bm k},p} \}$ indicates an $m$-tuple of bosonic operators with $p = 1,...,m$. The diagonalization of ${\cal H}_{\bm k}$ which preserves the bosonic commutation relation is achieved via a Bogoliubov transformation \cite{blaizot-ripka}, $ ( \{ \beta_{\bm k, r} \},\{ \beta^{\dagger}_{-\bm k, r} \} ) = T_{\bm k}  ( \{ b_{\bm k, p} \},\{ b^{\dagger}_{-\bm k, p} \} )$, introducing new bosonic operators  $\beta_{\bm k, r}, \beta^{\dagger}_{\bm k, r}$ such that the Hamiltonian takes the form
\begin{equation}
{\cal H}_2 = \sum_{\bm{k},r}{\omega_{\bm{k}}^{(r)}}\left(\beta^{\dagger}_{\bm k, r}\beta_{\bm k, r}+\dfrac{1}{2}\right),
\end{equation}
namely the form of a free-quasiparticle Hamiltonian with eigenfrequencies $\omega_{\bm k}^{(r)}$ organized in $m$ bands ($r = 1, ..., m$).
The Bogoliubov matrix $T_{\bm k}$ has the property that  $T_{\bm k} \eta T^{\dagger}_{\bm k} \eta = \mathbb{1}$ and $T_{\bm k} \eta {\cal H}_{\bm k} T^{-1}_{\bm k} = \Omega_{\bm k}$, where we have introduced the matrices
\begin{eqnarray}
\eta & = & \begin{pmatrix} \mathbb{1}_{m}  & \mathbb{0}_{m} \\ \mathbb{0}_{m}  &  -\mathbb{1}_{m} \end{pmatrix}  \nonumber \\
\Omega_{\bm k} &=& {\rm diag}(\omega^{(1)}_{\bm k},...,\omega^{(m)}_{\bm k},-\omega^{(1)}_{\bm k},...,-\omega^{(m)}_{\bm k})~.
\end{eqnarray}
It shall be useful for the following to mention explicitly that the $T_{\bm k}$ matrix has the structure 
\begin{equation}
T_{\bm k} = \begin{pmatrix} U_{\bm k} & V^{*}_{\bm k} \\ V_{\bm k} & U^{*}_{\bm k} \end{pmatrix}
\end{equation}
where the $U_{\bm k}$ and $V_{\bm k}$ matrices are composed of column vectors of length $m$ 
\begin{equation}
U_{\bm k} = \left ( {\bm u}^{(1)}_{\bm k}~ ....~ {\bm u}^{(m)}_{\bm k}  \right  ) ~~~~  V_{\bm k} = \left ( {\bm v}^{(1)}_{\bm k}~ ....~ {\bm v}^{(m)}_{\bm k}  \right  )
\end{equation}
with the property of $\eta$-orthogonality, $\left (\bm{u}_{\bm k}^{(r)}, \bm{v}_{\bm k}^{(r)} \right ) \eta \left ( \bm{u}_{\bm k}^{(r')}, \bm{v}_{\bm k}^{(r')} \right )^{\dagger} = \delta_{rr'}$. 
The ${\bm u}^{(r)}_{\bm k}$, ${\bm v}^{(r)}_{\bm k}$ vectors dictate the spatial structure on the unit cell belonging to the ${\bm k}$-vector eigenmode belonging to the $r$-th band. 

\begin{center}
\begin{figure}[t]
\includegraphics[width=\columnwidth]{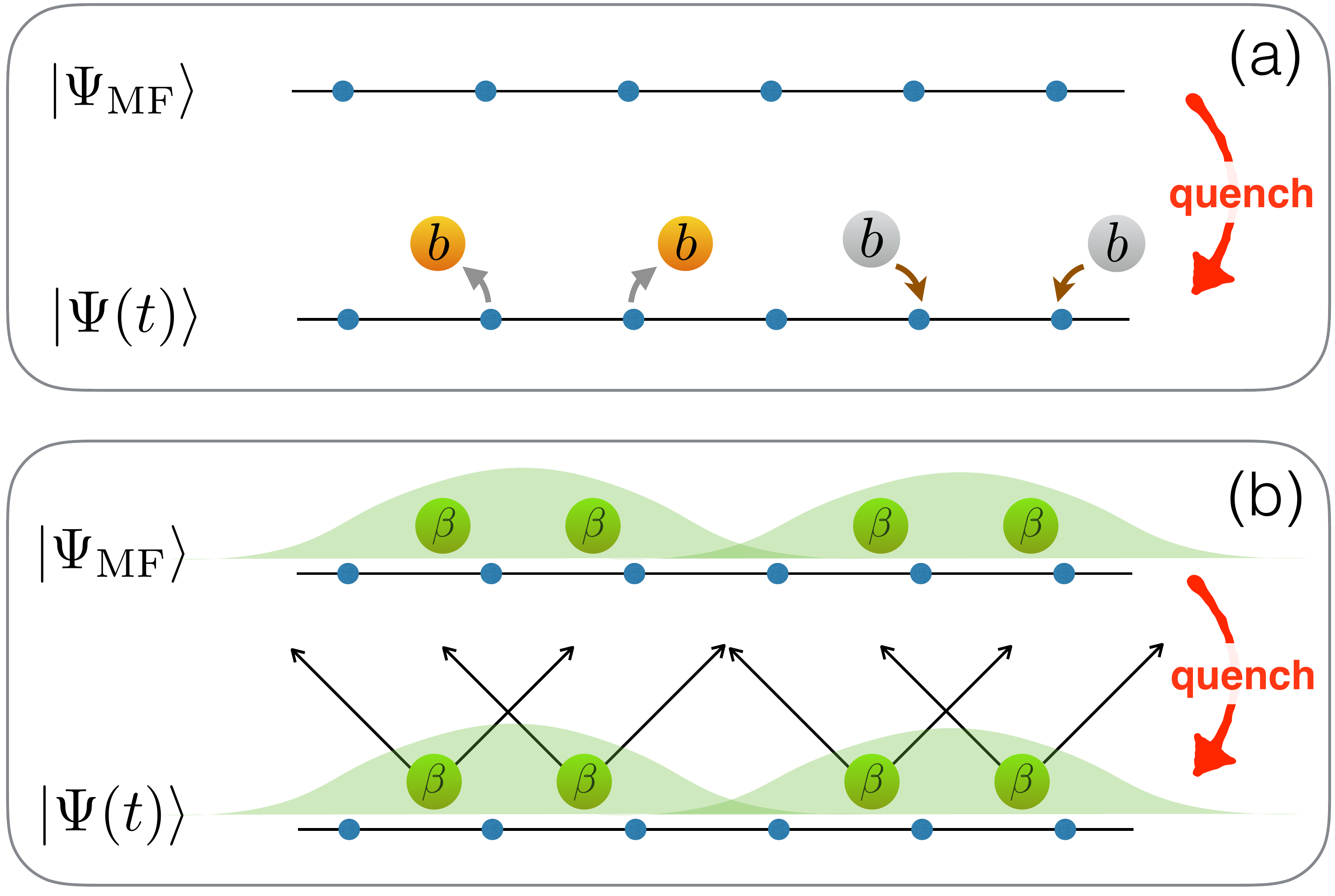}
\caption{Quantum quenches from mean-field states within the LSW picture. (a) $b$-boson picture: the initial state is the $b$ vacuum, and the evolution produces/annihilates pairs of $b$ bosons on neighboring sites; (b) $\beta$-boson picture: the initial state has a finite density of localized $\beta$ quasiparticles on each site, while the evolved state sees the quasiparticles expand ballistically.}   
\label{f.quench-sketch}
\end{figure}
\end{center}

\subsection{Quantum quenches within the linear spin-wave approach}
\label{quench_bbetabosons}

  The real-time dynamics of the quantum spin model of interest within LSW theory is reduced to the dynamics of coupled harmonic oscillators. Using the MF ground state as the initial state amounts to initialize the dynamics in the vacuum of $b$ bosons -- which is nonetheless a state containing a finite density of $\beta$ quasiparticles at each site. The points of view of the $b$ quasiparticles and $\beta$ quasiparticles on the ensuing dynamics are complementary and both useful. 
  
  In terms of $b$ bosons -- see Fig.~\ref{f.quench-sketch}(a) -- the quadratic Hamiltonian Eq.~\eqref{e.H2} contains pair-creation terms $b^{\dagger}_{lp} b^{\dagger}_{l'p'}$ (stemming from the $S'^x S'^x$ term in the transformed Hamiltonian, Eq.~\eqref{e.Ham'}) that will generate a finite density of quasiparticles from the initial vacuum. When starting from the MF ground state, the LSW approach is obviously very accurate at the beginning of the evolution; and it keeps its validity if, during the evolution, the local density of $b$ bosons remains sizably small, namely under the condition 
  \begin{equation}
   r(t) = \frac{1}{2Ns} \sum_{lp} \langle b_{lp}^{\dagger} b_{lp} \rangle(t) \ll 1
   \label{e.r}
  \end{equation} 
  where $\langle ... \rangle(t)$ represents the expectation value on the evolved state at time $t$.  Stated otherwise, if the $b$ bosons form a dilute gas at all times, the interactions among them can be safely neglected. This approximation will eventually break down at long times, because interactions among $b$ bosons are essential to describe the thermalization process of the system. Yet the existence of a (quasi-)stationary regime within LSW theory during which $r(t) \approx {\rm const.} \ll 1$ suggests that the TFIM of interest, despite being non-integrable and hence thermalizing to an ordinary Gibbs ensemble, may exhibit a phenomenon of prethermalization \cite{Langenetal2016} -- namely an initial relaxation towards a quasi-stationary state purely emerging from the dephasing of uncoupled modes. This should be valid for quenches sufficiently small for the LSW description to apply, namely under the condition that the injected energy produces a diluted gas of quasiparticle excitations \footnote{Prethermalization could offer an interpretation to the seeming failure of the square-lattice TFIM to thermalize under specific quench protocols, as reported by the numerical study in Ref.~\cite{BlassR2016} -- although a quantitative test of this interpretation is left for future work.}.   

From the alternative point of view of the $\beta$ bosons (Fig.~\ref{f.quench-sketch}(b)), the initial MF state has a finite density of quasiparticles, which form a coherent state of pairs \cite{blaizot-ripka}. Given the relationship
\begin{equation}
\begin{pmatrix} \{ b_{\bm k,p} \} \\ \{ b^{\dagger}_{-\bm k,p} \} \end{pmatrix} =
  \begin{pmatrix} U^{\dagger}_{\bm k} & -V^{T}_{\bm k} \\ -V^{\dagger}_{\bm k} & U^T_{\bm k} \end{pmatrix}
\begin{pmatrix} \{ \beta_{\bm k,r} \} \\ \{ \beta^{\dagger}_{-\bm k,r} \} \end{pmatrix}
\end{equation}
and the fact that the MF state is the vacuum of the $b$ bosons, one obtains
\begin{eqnarray}
|\Psi_{\rm MF}\rangle & = & {\cal N}~ \exp(-K)~ |0\rangle_\beta \nonumber \\
K & = &  \frac{1}{2} \sum_{\bm k} \sum_{rr'} \beta_{\bm k,r}^{\dagger} \left[ \left (U_{\bm k}^\dagger\right )^{-1} V_{\bm k}^{\dagger} \right ]_{rr'}  \beta_{\bm k,r'}^{\dagger} 
\end{eqnarray}
where $\cal N$ is a normalization factor. Here $|0\rangle_\beta$ is the vacuum of the $\beta$ quasiparticles, namely the LSW ground state. Hence the MF state can be viewed as a dilute gas of $\beta$-boson pairs, entangling eigenmodes of the quadratic Hamiltonian with opposite momenta within the same band and among different bands. Yet the real-space nature of this state remains extremely simple, as entanglement is absent altogether in real space, and $\exp(-K)$ is an operator which completely disentangles the entangled LSW ground state to map it onto the MF state.  Table~\ref{betabosons_populations} shows the populations of $\beta$ quasiparticles of the various bands in the MF ground states for the three models of interest in this work. 

\begin{table}
		\begin{tabular}{|c  c | c c c |}
		\toprule
		 Lattice model & $(\Gamma,H)$ & $n^{(1)}$ & $n^{(2)}$ & $n^{(3)}$  \\
		 \hline
	      triangular $(|\Psi_{\rm MF}^{(+)}\rangle)$ &(0.6,0.3) &  $10^{-2}$  & $2.6* 10^{-3}$ & $ 8.4 * 10^{-4}$ \\		
	      Lieb $(|\Psi_{\rm MF}^{(-)}\rangle)$  & (1,0) & $3.8*10^{-3}$  & $0$ & $ 2.1*10^{-3}$  \\
	      Kagom\'e $(|\Psi_{\rm MF}^{(-)}\rangle)$ & (1,0.) & $7.1*10^{-4}$  & $6.4*10^{-5}$ & $ 2.9*10^{-4}$ \\
		  \hline
		\end{tabular}
		\caption{ \label{betabosons_populations}  
	Quasiparticle densities $n^{(r)} = (1/V) \sum_{\bm k} \langle \beta^{\dagger}_{\bm k,r} \beta_{\bm k,r}\rangle$ calculated on the MF ground states $(|\Psi_{\rm MF}^{(\pm)}\rangle)$ of the lattice models of interest to this work.}
		\end{table}

 The quench dynamics leads then to the ballistic expansion of the $\beta$ quasiparticle pairs, with conservation of the average quasiparticle number. Such an expansion induces the spreading of correlations and of entanglement in real space, within a causal cone which is determined by the velocity spectrum of the quasiparticles themselves, and whose aperture is determined by the maximum group velocity from the quasiparticle dispersion relation - acting as an effective "speed of light" in the system \cite{calabrese_evolution_2005}.      

 From a technical point of view, initializing the quench dynamics from the vacuum of $b$ bosons suggests to use the Heisenberg picture in the calculation of time-evolved observables $\hat{A} \to \hat{A}(t) = \exp(i\hat{\cal H} t) \hat{A} \exp(-i\hat{\cal H} t)$, and to extract the evolved average values $\langle A \rangle(t)$ as vacuum expectation values of $\hat{A}(t)$ (suitably expressed as a function of the $b,b^{\dagger}$ operators).

 In the following we shall discuss the application of LSW theory to quench dynamics for the study of several lattices, each exhibiting one (nearly) flat band out of the various bands $\omega_{\bm k}^{(r)}$ into which the quasiparticle dispersion relation organizes.

 \begin{center}
 \begin{figure}[t]
 \includegraphics[width=0.6\columnwidth]{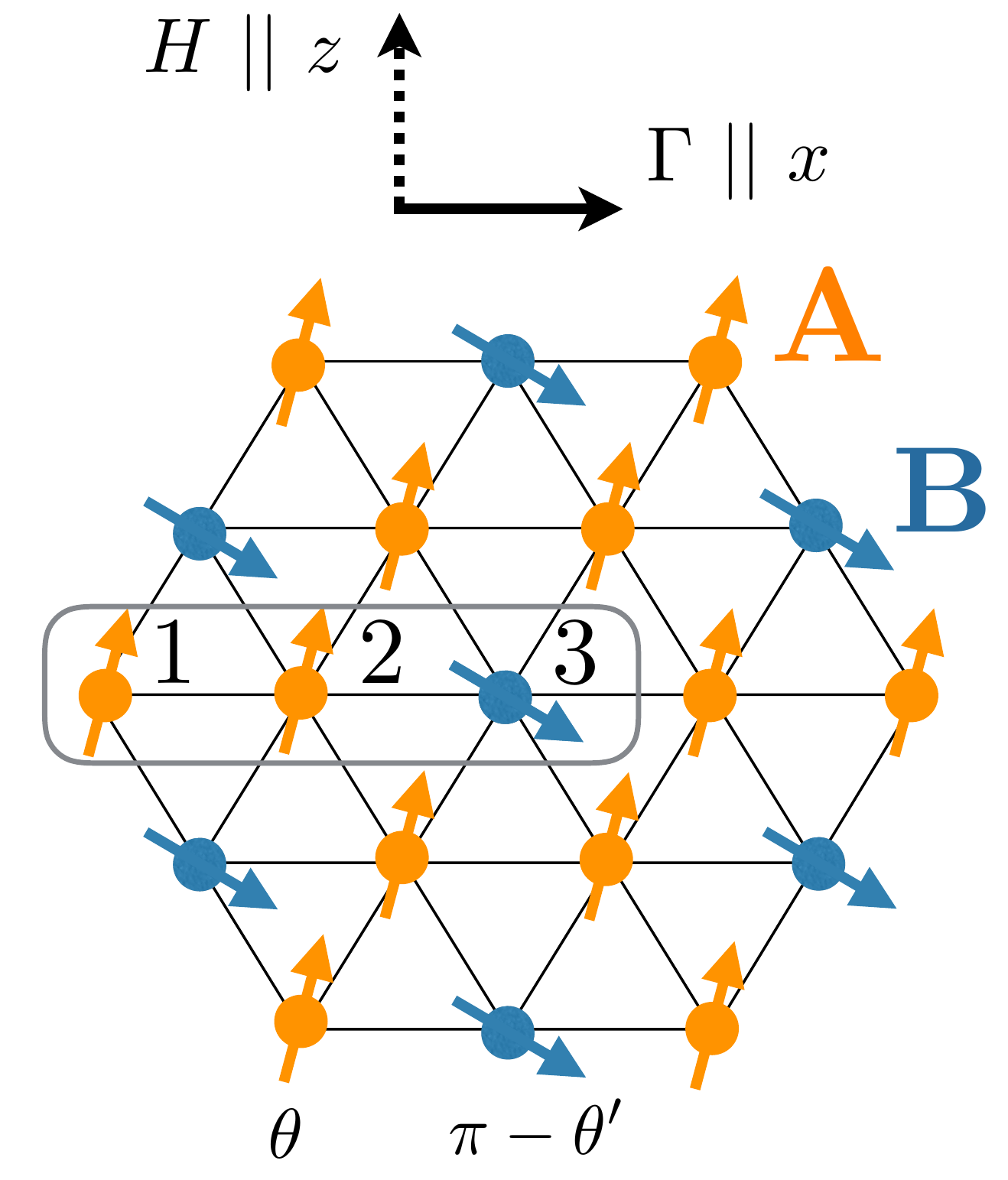}
 \caption{(Mean-field states on the triangular lattice: three-sublattice state in a transverse plus longitudinal field. The boxed area indicates the magnetic unit cell, and it shows the $p$-index convention which is used in the text.}
 \label{f.triangleMF}
 \end{figure}
 \end{center}

\section{Antiferromagnetic triangular lattice}
\label{s.triangular} 

\subsection{Ordered ground state in zero longitudinal field and linear spin-wave theory}

 We shall start our discussion from the TFIM on the triangular lattice model with nearest-neighbor antiferromagnetic interactions. There, as we shall see, flat-band physics is induced by frustration, namely by the intrinsic competitive nature of antiferromagnetic interactions on a non-bipartite lattice such as the triangular one. In the absence of any external field, $H = \Gamma = 0$, the minimization of the Hamiltonian Eq.~\eqref{e.Ham} with nearest-neighbor antiferromagnetic interactions is obtained locally by satisfying an ``up-down rule" of having at least two antiparallel spins on each triangular plaquette. Such a rule does not define a unique ground state (modulo a global spin flip) but rather a vastly degenerate manifold, growing exponentially with system size, as first shown by Wannier ~\cite{wannier_antiferromagnetism._1950}. The application of a transverse field $\Gamma$ lifts this exponential degeneracy in favor of a finitely degenerate ground state, which exhibits long-range order, giving rise to a paradigmatic example of the quantum order-by-disorder mechanism \cite{moessner_ising_2001}. The ordered state has a 3-site unit cell $AAB$ defined \emph{e.g.} on three horizontally adjacent sites (see Fig.~\ref{f.triangleMF}(a)), with a spin configuration of the kind $\nearrow\nearrow\searrow$ or $\searrow\searrow\nearrow$, which includes antiferromagnetism along the (vertical) $z$ axis and a tilt along the (horizontal) magnetic field axis.  This structure defines a unique tiling of the lattice obeying the up-down rule for the $z$ spin components. 
  
   Such an ordered ground state represents a potentially good starting point for the LSW treatment. Building the LSW approach around the order-by-disorder mechanism amounts to searching the MF ground state as a state with a 3-site unit cell, with three angles $\theta_1$, $\theta_2$ and $\theta_3$ providing the tilts with respect to the $z$ axis that the three sublattices experience under the application of the transverse field $\Gamma$. The minimization of the MF energy with respect to the angles provides a solution in the form $\theta_1 = \theta_2=\theta$  and $\theta_3 = \pi-\theta'$ for the three sublattices (see Fig.~\ref{f.triangleMF}(a) for a sketch). 
   
   Yet when building LSW theory around this MF state as described in the previous section, we systematically find normal modes with \emph{imaginary} frequencies $\omega_{\bm k}^{(r)}$, revealing that the harmonic approximation we are trying to build around a hypothetically stable minimum possesses in fact directions of instability. Such an instability is rather remarkable, witnessing that the order-by-disorder mechanism produces an ordered ground state whose quantum fluctuations are intrinsically anharmonic, and too strong to be captured faithfully within the LSW approach.   
   Interestingly the instability is present regardless of the value of $\Gamma$ below the critical value $1.5J$, at which the MF solution gets polarized by the transverse field.

  \begin{center}
 \begin{figure*}[t]
 \includegraphics[width=\textwidth]{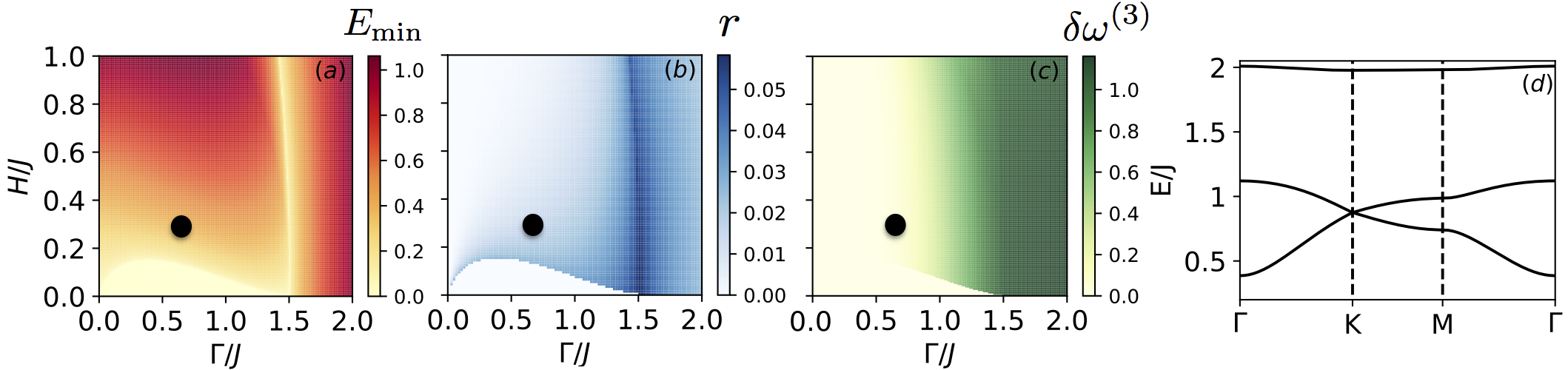}
 \caption{Linear spin-wave theory for the TFIM on the triangular lattice. (a) Lowest band gap of the triangle lattice spectrum as a function of the normalized parameters $(\Gamma/J,H/J)$. A vanishing of the gap in the low-$H$ region corresponds to the appearance of imaginary frequencies. (b) Ground-state quasiparticle density $r$. (c) Bandwidth of the upper band, $\delta\omega^{(3)}$. (d) Band  structure at the point $\Gamma/J=0.6,H/J=0.3$, indicated by a black dot on the other three panels.}
 \label{triangle_characteristics}
 \end{figure*}
 \end{center}

\subsection{Linear spin-wave theory in a finite longitudinal field and flat-band spectrum}
 
  \subsubsection{Dispersion relation in a longitudinal field}
  \label{s.dispersionH}
 
 The difficulty of LSW to deal with the triangular lattice TFIM can be circumvented by adding to the system a longitudinal field $H$, which, already in the absence of the transverse field, lifts the exponential degeneracy of the classical ground state to favor the 3-sublattice $\uparrow\uparrow\downarrow$ structure. Indeed the  up-down rule on each triangle, when supplemented with the requirement that a majority of the spins be in the $\uparrow$ configuration, defines a unique tiling of the triangular lattice, modulo two translations -- which leave a three-fold degenerate ground state only. Adding a transverse field to the model introduces quantum dynamics and allows for a meaningful LSW approach. 
 
 The MF ground state in the presence of a longitudinal plus transverse field is slightly modified with respect to its $H=0$ counterpart, in that the tilt angles $\theta$ and $\theta'$ caused on the three sublattices by the transverse field are slightly modified. When building LSW theory around the MF solution, the $H$ field is found to have the desired effect of removing the imaginary frequencies over an extended portion of the $(\Gamma,H)$ plane, as shown in Fig.~\ref{triangle_characteristics}(a). In that same portion of the phase diagram, the $r$ parameter of Eq.~\eqref{e.r}, controlling the validity of the LSW approximation, is appreciably small on the ground state; as shown in Fig.~\ref{triangle_characteristics}(b), it becomes sizable only at the boundaries of the LSW stability region and around the critical line between the 3-sublattice ordered state and the polarized state.
 
 Within the range of validity of LSW theory the excitation spectrum exhibits two low-lying bands touching at a Dirac cone, and a well-separated upper band which is nearly flat on the scale of the bandwidth of the other bands -- see Fig.~\ref{triangle_characteristics}(d). The nearly flat characteristic of the upper band persists over a large portion of the LSW stability region, as shown in Fig.~\ref{triangle_characteristics}(c). To understand the large difference in the bandwidths of the various band of the system, one can simply inspect the form of the corresponding eigenmodes at a given wavevector ${\bm k}$. In general we observe 
 \begin{eqnarray}
\text{lowest band} ~~~~~ \bm{u}^{(1)}_{\bm{k}}&=&(a^{(1)}_{\bm{k}}, -a^{(1)}_{\bm{k}}, b^{(1)}_{\bm{k}}) \nonumber \\
 \text{middle band} ~~~~~ \bm{u}^{(2)}_{\bm{k}} &=& (a^{(2)}_{\bm{k}},a^{(2)}_{\bm{k}},b^{(2)}_{\bm{k}}) \nonumber \\
 \text{highest band} ~~~~~  \bm{u}^{(3)}_{\bm{k}}&=& (a^{(3)}_{\bm{k}},a^{(3)}_{\bm{k}},b^{(3)}_{\bm{k}})
 \end{eqnarray}
 and similarly for the ${\bm v}_{\bm k}$ vectors.
 
 \begin{center}
\begin{figure}[t]
\includegraphics[width=0.8\columnwidth]{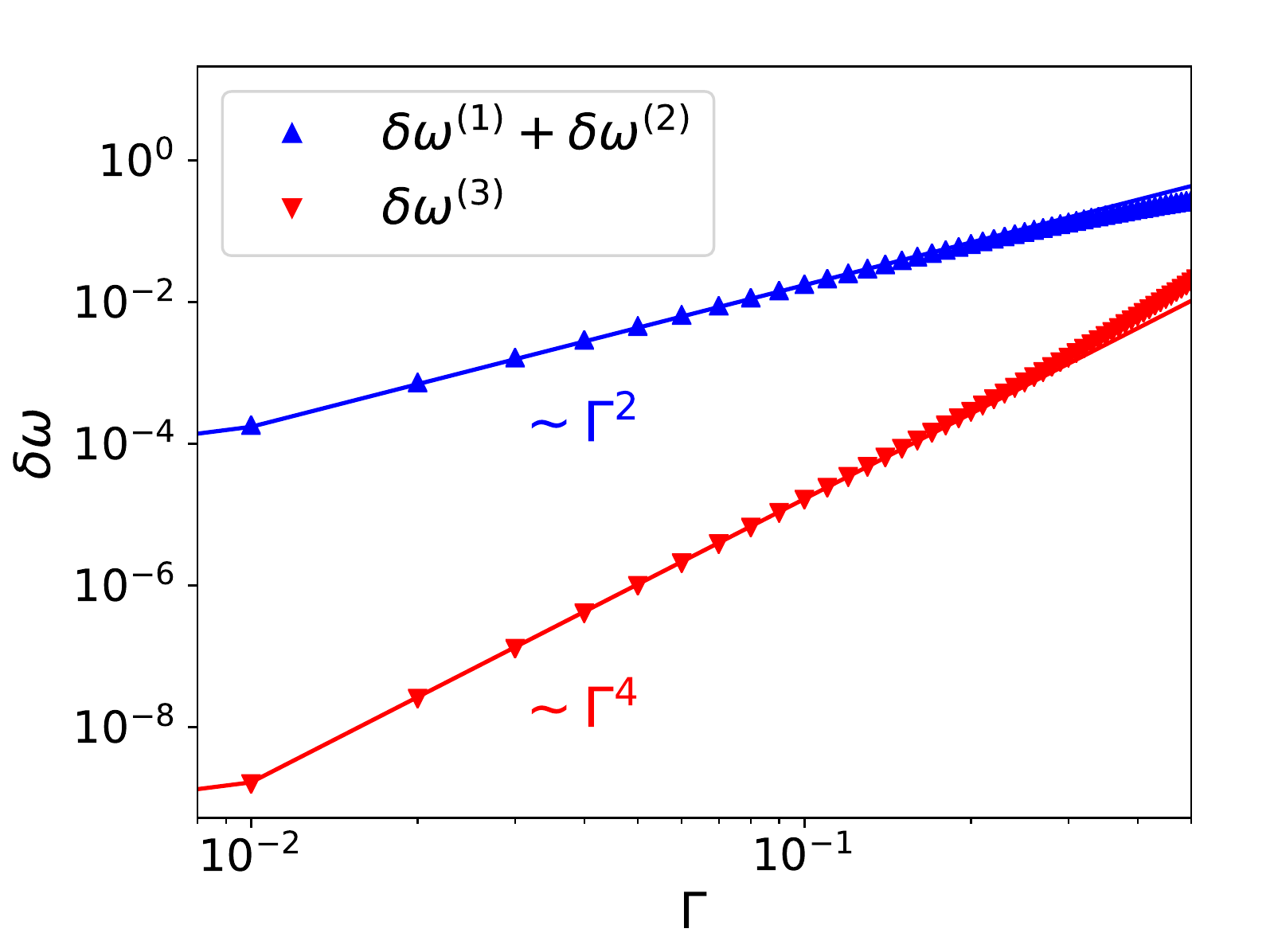}
\caption{Transverse-field dependence of the total bandwidth for the lower bands ($\omega^{(1)}, \omega^{(2)}$) and of the bandwidth of the upper band ($\omega^{(3)}$) for $J=1$ and $H=0.9$. The solid lines represent fits to a $\Gamma^{2}$ (upper) and $\Gamma^4$ behavior (lower).}
\label{band_opening}
\end{figure}
\end{center}
 
 Here the three amplitudes for each mode are referred to the sites in the unit cell, according to the convention of Fig.~\ref{f.triangleMF}. The fact that amplitudes are equal on the sites $p=1$ and $2$ is a consequence of the equivalent MF spin configuration that the two sites host. In particular we observe that $\vert a^{(1,2)}_{\bm{k}}\vert\gg\vert b^{(1,2)}_{\bm{k}}\vert$, namely the modes of the two lowest bands have support predominantly on the $A$ sites (see Fig.~\ref{f.triangleMF}); the distinction between the lowest and middle band resides in the fact that the two $A$ sites oscillate in phase opposition and in phase, respectively.  On the other hand, the highest band has support mainly on the $B$ sites, $\vert a^{(3)}_{\bm{k}}\vert\ll\vert b^{(3)}_{\bm{k}}\vert$. As the $B$ sites are not connected to each other directly, the propagation of excitations which are confined to the $B$ sublattice goes necessarily through an intermediate hopping event onto the $A$ sublattice. As we shall see shortly, this is a highly off-resonant mechanism, suggesting that the effective hopping of excitations between $B$ sites passing via the $A$ sites can be much smaller than the direct hopping of excitations between $A$ sites; and leading to the vast difference in bandwiths among the bands.

\subsubsection{Perturbative treatment of the transverse field}

 To understand why the hopping of excitations from the $A$ to the $B$ sublattice is strongly non-resonant, we can take the limit $J, H \gg \Gamma$ and treat the transverse field - responsible for the hopping of excitations - perturbatively.  
 In the limit $\Gamma = 0$ elementary excitations correspond to localized spin flips: starting from the $AAB = \uparrow\uparrow\downarrow$ configuration, the flip of an $A$ spin costs an energy $H$, uniquely given by the field term. Indeed the $A$ sites are fully flippable for what concerns the spin-spin interactions, being surrounded by as many $\uparrow$ as $\downarrow$ spins. On the other hand the $B$ sites are not flippable for what concerns the interaction, and indeed their flip costs an energy $6J -H$. Under the condition $H \ll 3J$ there is therefore a significant energy mismatch $\Delta = E_B - E_A =  2(3J-H)$ between a spin flip localized on the $ B$ sublattice and one localized on the $A$ sublattice.     

  The introduction of a small transverse field leads to a tilt of the local spin configuration along the $x$ axis, so that excitations are now defined as spin flips with respect to a locally tilted quantization axis $z'$ (and generated by the $b^{\dagger}$ operator). The 
hopping of spin flips from one site to the next is mediated by the $S'^x S'^x$ term in the transformed Hamiltonian Eq~\eqref{e.Ham'}, which in the low-$\Gamma$ limit has amplitude $J \sin^2\theta \sim \Gamma^2/J$, given that $\sin\theta\approx \Gamma/\Gamma_c$ where $\Gamma_c = (3/2) J$ (for $H=0$, and slightly lower for finite $H$, see Fig.~\ref{triangle_characteristics}). 
  Therefore we expect that spin flips created on an $A$ site at an energy cost $H + {\cal O}(\Gamma)$ propagate resonantly on the $A$ sublattice - which forms a fully connected honeycomb lattice - acquiring a dispersion relation with a bandwidth $w^{(1,2)} \sim \Gamma^2/J$ for the lowest and middle band -- and this is indeed the correct low-$\Gamma$ scaling for the bandwidth of these bands, as shown in Fig.~\ref{band_opening}. Parenthetically we also notice that the Dirac cone exhibited by the two lowest subbands is directly related to the effective honeycomb geometry of the associated modes, and indeed it sits at the $K$ point in the magnetic Brillouin zone (which is the same as the one of the $A$ honeycomb lattice). 
  
  On the other hand, a spin flip created on a $B$ site sits at a much higher energy $6J -H  + {\cal O}(\Gamma)$, and, in order to propagate to a resonant $B$ site, it must perform a virtual transition through a low-energy, off-resonant $A$ site, leading to an effective hopping $\sim \Gamma^4/J^3$. This effective hopping dictates the scaling of the bandwidth $w^{(3)}$ for the highest band in the low-$\Gamma$ regime. Such a scaling is indeed observed in the exact solution shown in Fig.~\ref{band_opening}, explaining the vast difference in widths among the bands.

\begin{figure*}[t]
\centering
  \includegraphics[scale=0.5]{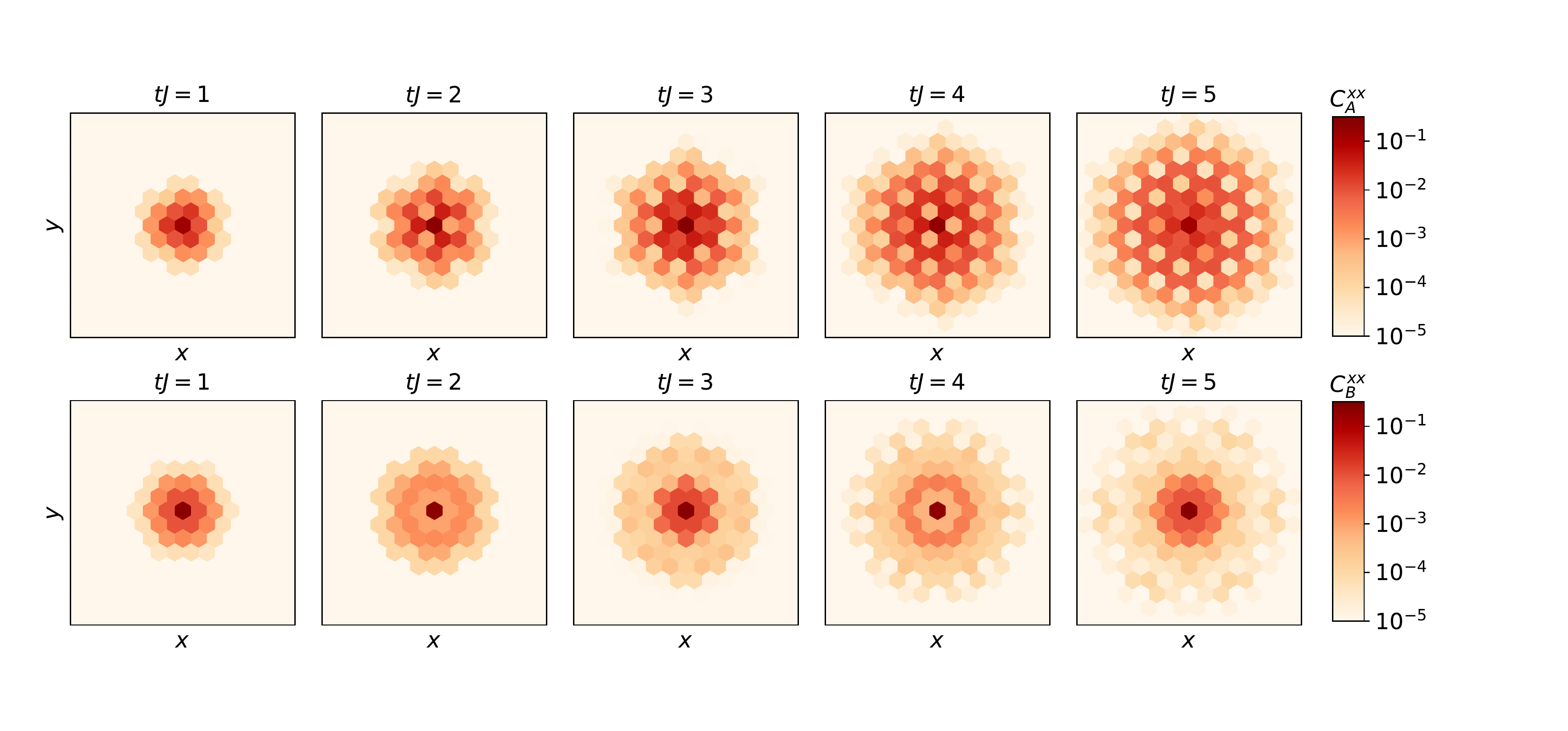}
  \caption{Propagation of correlations in the triangular lattice with parameters $(\Gamma/J=0.6,H/J=0.3)$ after quenching from the mean-field state. The first row shows the correlations between a site in the $A$ sublattice ($p=2$) and the rest of the lattice, while the second row represents the same quantity referred to a site in the $B$ sublattice.}
  \label{percolation_correlation}
\end{figure*}


\subsection{Quench dynamics}

 The very rich structure of the excitation spectrum revealed in the previous section can be fully reconstructed by investigating the dynamics of correlations after global and local quantum quenches. All the results presented in the following refer to the particular choice of parameters $H = 0.3 J$, $\Gamma = 0.6 J$, for which the highest band is nearly flat compared to the two lower ones $(w^{(3)}/(w^{(1)}+w^{(2)}) \approx 0.04)$. 

\subsubsection{Spreading of correlations after a global quench}

 Let us first focus on the quench protocol in which the dynamics is triggered by initializing the system in the MF ground state. We shall then explore how the propagation of harmonic excitations captured by LSW theory induces the onset of correlations (completely absent in the initial state); and how the dynamics of correlations is sensitive to the existence of excitations modes with vastly different dispersion laws. 
In particular we focus on the spin-spin correlation function for the $S^x$ spin components, namely 
\begin{equation}
\label{e.corr}
    C^{xx}\left({\bm r}_{lp},{\bm r}_{l'p'};t \right)=\langle\delta S^{x}_{lp} \delta S^{x}_{l'p'}\rangle (t)
\end{equation}
which turns out to be the strongest form of correlation developed by the dynamics. The system is translationally invariant modulo the $AAB$ unit cell, so that the correlation functions to monitor are essentially of two types: 1) $C^{xx}_A(\bm r' - \bm r)$ when taking the reference site at position ${\bm r}$ on the $A$ site labeled \emph{e.g.} by $p=2$ of the unit cell (as done in Fig.~\ref{percolation_correlation} -- the correlation function starting from the $p=1$ site is obtained from $C^{xx}_A$ by reflecting its argument around the vertical axis bisecting the $AA$ bond); and 2) $C^{xx}_B(\bm r' - \bm r)$, obtained by taking the reference site on the $B$ sublattice. 
 
 Fig.~\ref{percolation_correlation} shows the post-quench time evolution of the $C^{xx}_A$ and $C^{xx}_B$ correlations. For both correlation functions one clearly observes the light-cone effect, namely correlations at time $t$ have only appeared up to a distance $r\sim t/v_{\rm LR}$, where $v_{\rm LR}$ is a characteristic (Lieb-Robinson) velocity, given by $2 \max_{{\bm k},r} \vert\bm{v}^{(r)}_G(\bm k)\vert$ in the case of well-defined quasiparticle excitations at hand, where ${\bm v}^{(r)}_G(\bm k) = {\bm \nabla}_{\bm k} \omega_{\bm k}^{(r)}$ is the group velocity of the $r$-th band. 
 
  Yet the most striking feature of the spreading of correlations is offered by the internal structure of the light cone, revealing the spatial structure of the excitation modes. Indeed one sees that the correlations appearing in the system at early times are only established between $A$ sites - as revealed by $C_A^{xx}$, which develops a characteristic honeycomb pattern of correlations; or between $B$ and $A$ sites -- as revealed by $C_B^{xx}$, which develops a similar, yet much weaker honeycomb pattern. Hence one can conclude that the spreading of correlations at early times is carried by the propagation of the excitation modes mostly confined to the $A$ honeycomb lattice; and with lesser support on the $B$ sites, which nonetheless allows them to correlate weakly the latter sites with the $A$ sites. These are indeed the fast modes residing in the lowest bands. 
 
 The strong asymmetry between correlations among $A$ sites (or $AA$ correlations) and correlations among $B$ sites (or $BB$ correlations) at early times may suggest that the $BB$ correlations only establish at much later times, as they are carried by the very slow quasiparticles residing in the highest, nearly flat band. 
  But there is a further ingredient that we have neglected so far in our analysis, namely the specific structure of the quasiparticle populations $\langle \Psi_{\rm MF} | \beta^{\dagger}_{{\bm k},r} \beta_{{\bm k},r} | \Psi_{\rm MF} \rangle$ induced by the quench (namely calculated on the initial state -- see Sec.~\ref{quench_bbetabosons}) and conserved during the dynamics. Table~\ref{betabosons_populations} shows that the band populations associated with the MF ground state are very unequal, and that the highest band is weakly populated. This in turn explains the weakness of $BB$ correlations, which lack their privileged carriers (the quasiparticles in the highest bands) in the dynamics.

\begin{figure*}[t]
\centering
 \includegraphics[width=0.85\textwidth]{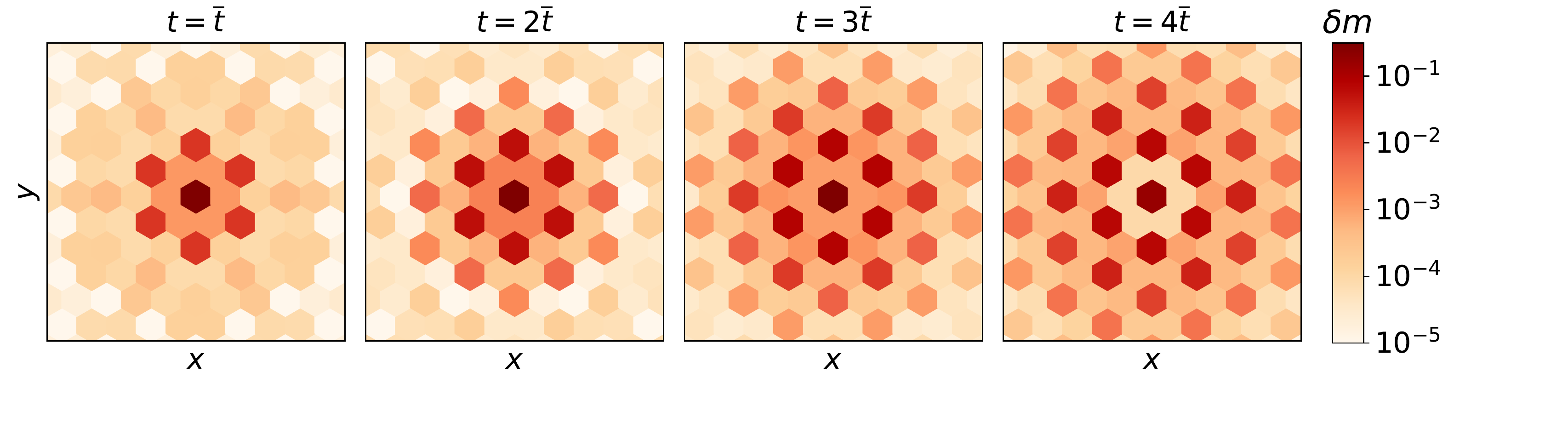}
    \caption{Spin deviation profile $\delta m_{lp}$ after a local/global quench starting from the MF ground state plus a spin flip on a site on the $B$ sublattice. The snapshots refer to integer multiples of the characteristic propagation time $\overline{t}=d/(2v^{(3)}_{G, \rm max})\simeq 50 J^{-1}$, where $d$ is the distance between two $B$ sites and $v^{(3)}_{G,\rm max}$ is the maximum group velocity in the upper band.}
    \label{slow_lightcone}
\end{figure*}

  \subsubsection{Global/local quench and flat-band excitations}
 
 To explicitly involve the flat-band quasiparticles in the dynamics one can proceed to a different quench protocol, namely a mixed global/local quench that initializes the system in the MF state with a spin flip at the $B$ site ($p=3$) of the $l_0$-th unit cell, $ |\Psi'_{\rm MF} \rangle = b^{\dagger}_{l_0,3} |\Psi_{\rm MF}\rangle$. Given that the flat-band excitations have maximum overlap with the $B$ sites, such a local flip creates a localized wavepacket of $\beta$ quasiparticles belonging to the flat band. From the technical point of view, the calculation of expectation values $\langle A \rangle (t)$ starting from this modified initial state amounts to the calculation of vacuum expectation values of the operator $b_{l_0,3} \hat{A}(t) b^{\dagger}_{l_0,3}$.
 
 The quench dynamics is monitored via the spreading of the initial spin deviation imposed on the MF ground state, namely by studying the spin-deviation profile with respect to the local quantization axis
 \begin{eqnarray}
 \delta m_{lp}(t) &=& 
 -  \langle \Psi'_{\rm MF} |  S'^z_{lp}(t)   |\Psi'_{\rm MF} \rangle  +  \langle \Psi_{\rm MF} |   S'^z_{lp}(t)  |\Psi_{\rm MF} \rangle \\
 &=& \langle \Psi'_{\rm MF} | \left(b^{\dagger}_{lp} b_{lp} \right)_t   |\Psi'_{\rm MF} \rangle -   \langle \Psi_{\rm MF} |   \left ( b^{\dagger}_{lp} b_{lp} \right)_t |\Psi_{\rm MF} \rangle \nonumber
 \label{e.spindeviation}
  \end{eqnarray}
 which is obtained by subtracting the evolved $b$-boson density profiles obtained by quenching from the MF ground state with and without the local spin flip.
 
 The spin deviation profile is shown in Fig.~\ref{slow_lightcone}. Its expansion is ballistic, as expected from a model of free quasiparticles. But, compared to the dynamics of correlations discussed in the previous subsection, the expansion dynamics 1) is highly non-uniform, taking place predominantly on the $B$ sites and 2) it is very slow, taking place over a characteristic timescale $\bar{t} = d/(2 v^{(3)}_{G,{\rm max}}) \sim (J/\Gamma)^4 J^{-1} \gg J^{-1}$ (for $\Gamma/J \ll 1$), where $v^{(3)}_{G,{\rm max}}$ is the maximum group velocity on the highest band and $d$ the lattice spacing. Hence this result directly reveals the existence of very slow excitation modes that are confined to the $B$ sublattice. By contrast, the amount of spin deviation which leaks on the $A$ sites is much weaker and expands much faster, witnessing that the initially localized spin flip also triggers fast modes, that nonetheless have very little support on the $B$ sites.    

\begin{center}
\begin{figure*}[t]
    \includegraphics[width=0.7\textwidth]{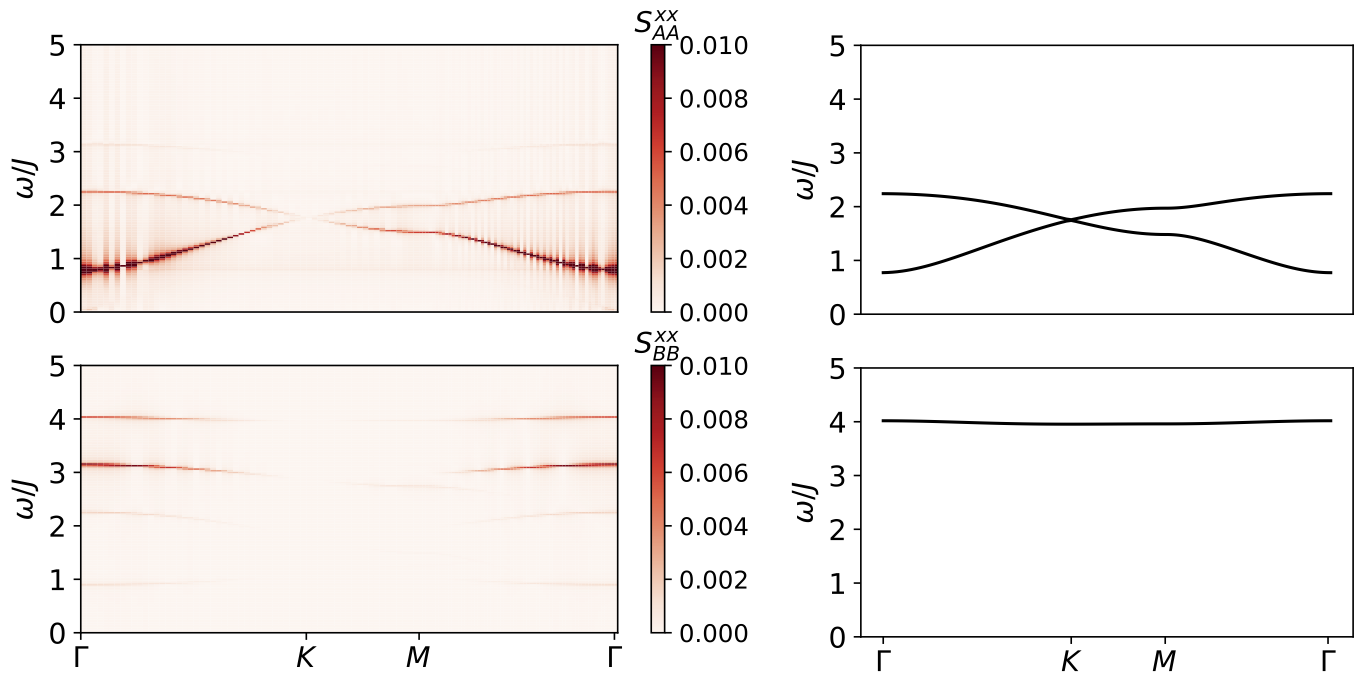}
    \caption{Quench spectroscopy for the triangular lattice TFIM with $\Gamma/J = 0.6$, $H/J = 0.3$ . Left side: power spectrum $\vert\mathrm{FT}[S^{xx}_{pp'}\left(\bm{k},t\right)](\omega)\vert^2$ computed numerically after a post-quench evolution over a time $200 J^{-1}$, with $p=p'$ on the $A$ sublattice (upper panel) and on the $B$ sublattice (lower panel). 
    Right side: Band structure $2 \omega^{(r)}_{\bm{k}}$ with $r=1,2$ (upper panel) and $r=3$ (lower panel).}
    \label{quench_spectroscopy}
\end{figure*}
\end{center}

  \subsubsection{Quench spectroscopy reconstructing the flat band}

 The quench protocols described above have the ability to reveal the existence of slow vs. fast modes, and they allow for an estimate of the order of magnitude of their bandwidth $w$  -- which, in the absence of singularities of the dispersion relation, controls the characteristic group velocity $ v_G \sim w/Q$ where $Q$ is the length of a basis vector in the reciprocal lattice. For a more quantitative reconstruction of the excitation spectrum, one can rely on the intuitive property that the eigenfrequencies of the evolution Hamiltonian must control the Fourier spectrum of the time evolution of observables, as $\langle A(t) \rangle = \sum_{\phi,\phi'} \langle \Psi_{\rm MF}|\phi\rangle \langle \phi' | \Psi_{\rm MF} \rangle \langle \phi | A | \phi' \rangle e^{i(\omega_\phi-\omega_{\phi '})t}$ with $\phi$, $\phi'$ labeling the Hamiltonian eigenstates. More specifically, in order to reconstruct the eigenfrequencies at a particular wavevector ${\bm k}$, one can focus on a ${\bm k}$-dependent observable: a very convenient one is the time-evolved structure factor defined \emph{e.g.} for the $S^x$ spin components as 
\begin{equation}
\label{fourier_convention}
 S^{xx}_{pp'}\left(\bm{k},t\right)=\sqrt{\frac{m}{N}} \sum_{l,l'} e^{i\bm{k}\cdot\left(\bm{r}_l-\bm{r}_{l'}\right)} ~C^{xx}\left(\bm{r}_{lp},\bm{r}_{l'p'};t \right)
 \end{equation}
 where ${\bm r}_l$ is the position of the unit cell (\emph{e.g.} the position of the $p=1$ site in the unit cell, ${\bm r}_l = {\bm r}_{l,1}$). The $S^{xx}_{pp'}$ structure factor probes selectively the spatial Fourier transform of the correlation function for $AA$ correlations, $BB$ correlations or $AB$ correlations, which, as we shall see, can provide precious information on the spatial structure of the excitation modes on the unit cell. 
 
  To gain insight into the excitation spectrum at wavevector ${\bm k}$ it suffices to observe that, within LSW theory \cite{Frerotetal2018}
  \begin{align}
    \label{corr_kspace}
    S^{xx}_{pp'}\left(\bm{k},t\right)& \simeq f_{\bm{k},pp'}+\sum_{r'\geqslant r} \Big \{ g^{rr'}_{pp'}({\bm{k}})\cos\left[\left(\omega_{\bm{k}}^{(r)}+\omega_{\bm{k}}^{(r')}\right)t\right]\notag\\
    &+i\overline{g}^{rr'}_{pp'}({\bm{k}})\sin\left[\left(\omega_{\bm{k}}^{(r)}+\omega_{\bm{k}}^{(r')}\right)t\right]\notag\\
    &+\sum_{r'<r} h^{rr'}_{pp'}({\bm{k}})\cos\left[\left(\omega_{\bm{k}}^{(r)}-\omega_{\bm{k}}^{(r')}\right)t\right]\notag\\
    &+i\overline{h}^{rr'}_{pp'}({\bm{k}})\sin\left[\left(\omega_{\bm{k}}^{(r)}-\omega_{\bm{k}}^{(r')}\right)t\right] \Big \}~.
\end{align}
The above result stems from retaining only the quadratic terms in the $b, b^\dagger$ operators contained in the full expression of the $C^{xx}$ correlation function -- see Appendix~\ref{app2}. Eq.~\eqref{corr_kspace} tells us that the Fourier spectrum of the time-evolved structure factor at wavevector ${\bm k}$ is uniquely composed by sums and differences of eigenfrequencies $\omega_{\bm k}^{(r)}$ corresponding to the various bands; and the spectral amplitudes of the various Fourier components $g^{rr'}_{pp'}$, $\bar{g}^{rr'}_{pp'}$ (for the frequencies $\omega_{\bm{k}}^{(r)}+\omega_{\bm{k}}^{(r')}$) and  $h^{rr'}_{pp'}$, $\bar{h}^{rr'}_{pp'}$ (for the frequencies $\omega_{\bm{k}}^{(r)}-\omega_{\bm{k}}^{(r')}$) -- whose explicit expressions are given in Appendix~\ref{app2} -- express the overlap of the modes of interest with the sites $p$ and $p'$ in the unit cell. This means that a Fourier analysis of the time dependence of $S^{xx}_{pp'}\left(\bm{k},t\right)$ gives access to the full structure of the dispersion relation; and, via the spectral weights, it also provides information on the unit-cell structure of the corresponding modes. This spectroscopic analysis of post-quench evolutions -- or \emph{quench spectroscopy} in short -- relies on the unique ability of quantum simulators to follow the dynamics of the system in real space and real time \cite{Labuhnetal2016}; it has already been successfully applied in the recent past \cite{Hungetal2013,Schemmeretal2017}; and it represents an interesting and powerful alternative to traditional spectroscopic techniques for bulk systems, based on scattering of external probes (such as light, neutrons, etc.). 

 The Fourier spectrum of $S^{xx}_{pp'}\left(\bm{k},t\right)$ obtained within the global quench protocol, by numerical Fourier transform of the post-quench evolution over the time window $tJ=200$, is shown in Fig.~\ref{quench_spectroscopy}. There we observe that the spectrum of the $S^{xx}_{AA}$ structure factor reveals the dispersive modes associated with the two lowest bands - namely it is strongly peaked at the frequencies $2\omega_{\bm k}^{(1)}$ and $2\omega_{\bm k}^{(2)}$. On the other hand, the Fourier transform of $S^{xx}_{BB}$ shows clearly the flat band (namely a peak at the frequency $2\omega_{\bm k}^{(3)}$), thanks to the strong overlap of the flat-band modes with the $B$ sites. 
 
\subsection{Connection to Rydberg quantum simulators}

 All the physical ingredients that we have invoked so far (from the model Hamiltonian, to the initial state preparation, and to the diagnostics of the evolved state) are perfectly compatible with the current capabilities of state-of-the-art experiments with Rydberg quantum simulators, based on arrays of individually trapped and addressable atoms \cite{Browaeysetal2016,Labuhnetal2016, barredo_atom-by-atom_2016}. The physics of the triangular lattice quantum Ising antiferromagnet has been probed in a recent experiment \cite{lienhard_observing_2017}, which attempted at the adiabatic preparation of the 3-sublattice ground state in a longitudinal plus transverse field (although the correlation pattern expected theoretically for the ground state has not been experimentally observed yet). Our work implicitly proposes a different scheme, in which the atoms are ``parachuted" to a low-energy state, namely they are prepared individually in a state which is chosen to be the best approximation to the actual ground state in a factorized (namely mean-field) form. This can be experimentally achieved by exploiting the individual addressability of the atoms and performing site-dependent Rabi pulses. Indeed, one can use the possibility of individually shaping the trapping potential seen by each atom, to differently light-shift the atomic levels on $A$ sites with respect to the $B$ sites. In so doing, a global Rabi pulse applied to atoms initially in the atomic ground (\emph{e.g.} $|\uparrow\rangle$) state, can rotate the atomic state on each sublattice differently, thereby achieving the desired periodic $AAB$ pattern. The subsequent non-equilibrium evolution is governed by the intrinsic Hamiltonian of the Rydberg atoms plus an external radiation field creating the $\Gamma$ term. The buildup of correlations under such an evolution has been already explored successfully in recent experiments \cite{Labuhnetal2016}.

\begin{center}
\begin{figure}[b]
\includegraphics[width=\columnwidth]{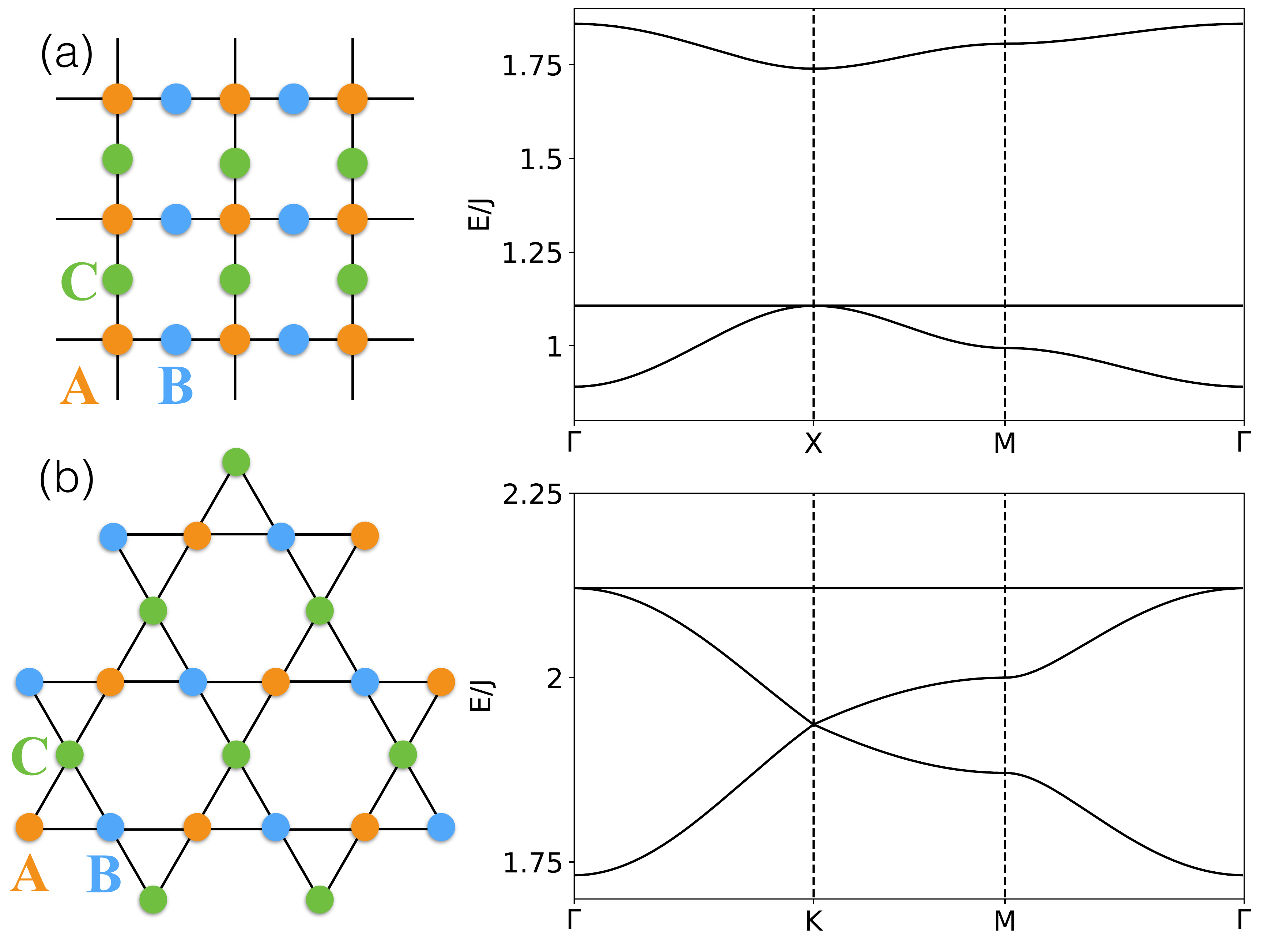}
\caption{Lattice structure and band structure of the Lieb and Kagom\'e lattices. (a) Magnetic sublattice structure and band structure for the ferromagnetic TFIM on the Lieb lattice with $\Gamma/J=1$. (b)  Magnetic sublattice structure and band structure for the ferromagnetic TFIM on the Kagom\'e lattice with  $\Gamma/J=1$ as well.}
\label{Lieb_kagome}
\end{figure}
\end{center}


\section{Kagom\'e and Lieb lattice}
\label{s.LiebKagome} 

 In this section we explore an alternative mechanism to the formation of flat bands, namely the appearance of localized eigenmodes of the Hamiltonian which cannot spread due to perfect destructive interference (Aharonov-Bohm caging). We shall focus on two well-known examples of lattices supporting such modes in the case of tight-binding Hamiltonians, namely the Lieb lattice and the Kagom\'e lattice. We shall show that flat bands occur as well in the dispersion relation of the elementary excitations of quantum Ising models cast on the same lattice, and that quench protocols allow for a direct inspection into the existence of localized spin waves. 
  
 In the following we shall focus on quenches starting from a \emph{ferromagnetic} initial state. The simplest such state, which has been used to initialize the dynamics in recent experiments \cite{Labuhnetal2016}, is the polarized state $|{\rm P}\rangle = |\uparrow\uparrow... \uparrow\rangle$. Obviously this state represents a highly excited state for the intrinsically antiferromagnetic Rydberg Hamiltonian ${\cal H}$ of Eq.~\eqref{e.Ham}, but, as we already pointed out in Sec.~\ref{s.MF}, the evolution of expectation values under the non-equilibrium dynamics starting from this state is identical when governed by ${\cal H}$ or by $-{\cal H}$; namely the evolution can be pictured as stemming from the low-energy dynamics of the ferromagnetic Hamiltonian $-{\cal H}$. Under this lens the LSW approach appears most suited to the task of studying the quench dynamics. From a technical standpoint, it is much more convenient for LSW theory to address the dynamics starting from a state which is a slight deformation of 
  $|{\rm P}\rangle$, namely from the MF ground state of ${-\cal H}$, $|\Psi_{\rm MF}^{(-)}\rangle$, in which the spins are tilted with respect to the $z$ axis by the transverse field. This is the setting under which the following results have been obtained. We expect the dynamics starting from the $|{\rm P}\rangle$ state not to differ significantly from the one studied here. Throughout the following discussion we shall assume that $H=0$, corresponding to zero detuning in the Rydberg-atom implementation.

 \begin{center}
\begin{figure*}[t]
\includegraphics[width=0.7\textwidth]{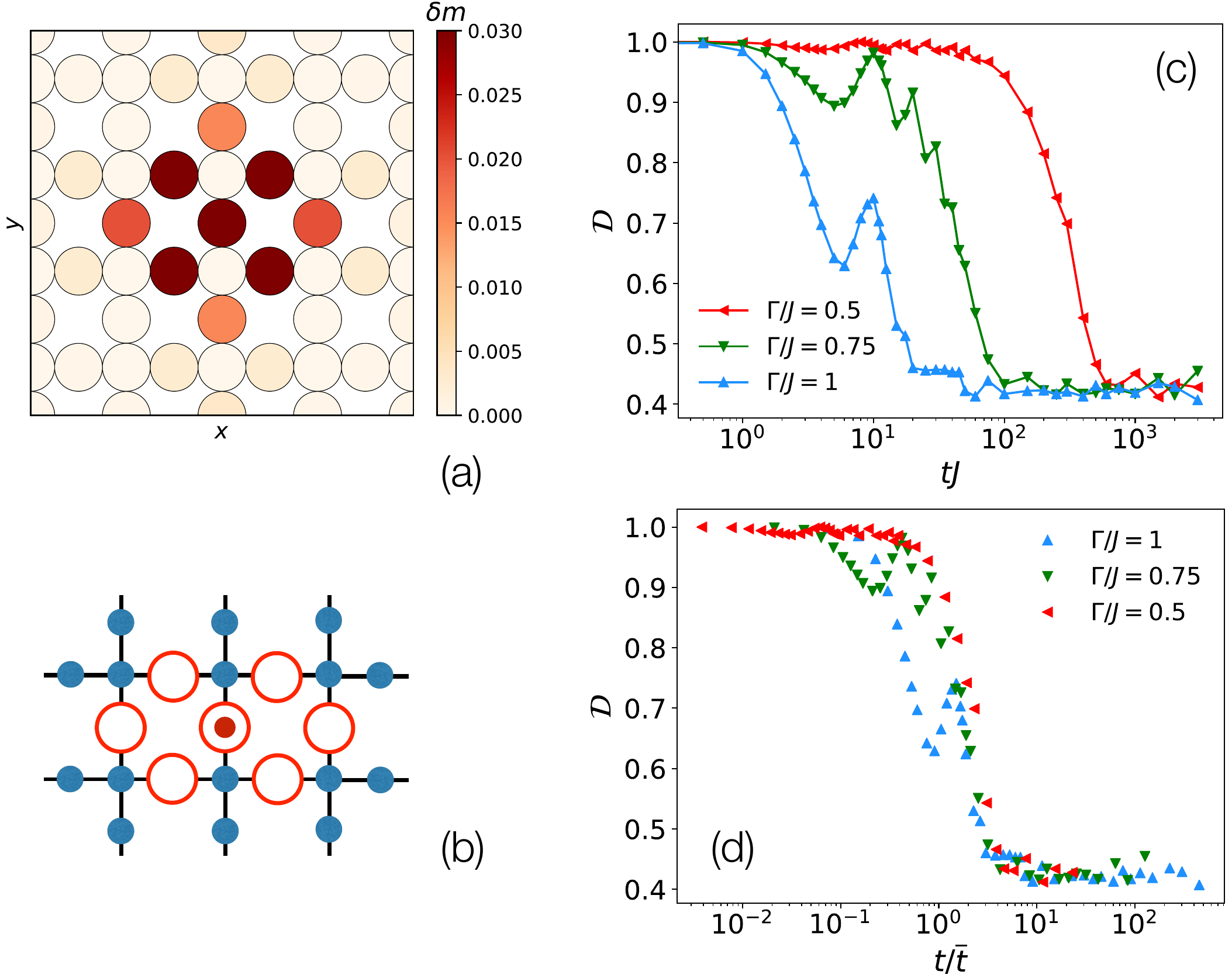}
\caption{Local/global quench on the Lieb lattice. (a) Spin deviation profile $\delta m_{lp}$ at long times following a local/global quench containing a spin flip of a link spin. (b) Open dots indicate the region LM$(i_0)$ covered by the maximally localized modes which overlap with the site $i_0$ (indicated by a red dot).  (c) Evolution of the localized fraction of the spin deviation computed for a $30\times 30$ Lieb lattice for different values of $\Gamma/J$. (d) Same quantity as in panel (c), plotted as a function of the characteristic time $\bar t$ (see main text).}
\label{local_quench_Lieb}
\end{figure*}
\end{center}

\subsection{Lieb lattice: global/local quench}
  
 The MF ground state of ${-\cal H}$ on the Lieb lattice exhibits two different tilting angles $\theta$ and $\theta'$ corresponding to sites at the nodes (hub sites) and on the bonds (link sites) of the square lattice, respectively.  Building LSW theory around this MF state, one obtains the excitation spectrum shown in Fig.~\ref{Lieb_kagome}(a), which exhibits a flat band in the center of the spectrum similarly to what is found for the tight-binding Hamiltonian on the same lattice \cite{Morales-InostrozaV2016}. 
Yet there are some relevant differences. In the tight-binding model the highest and lowest band, which are dispersive, touch each other as well as the flat band at the tip of a Dirac cone, and the spectrum is otherwise particle-hole symmetric. In the LSW spectrum, instead, the Dirac point is gapped out,
as the site-dependent tilting angles produce a corresponding onsite potential (through the last term of Eq.~\eqref{e.Ham'}). Moreover the flat band touches the lowest band at a point where the latter has a quadratic dispersion. 
 
 The diagonalization of the quadratic Hamiltonian using the ${\bm k}$ quantum number (as described in Sec.~\ref{LSW}) naturally assigns an extended nature to the flat-band modes, which nonetheless can be though of as superpositions of an extensive number of degenerate localized modes. In the following we shall refer to such superpositions as ${\bm k}$-modes. 
  Analyzing their  spatial structure, in a way similar to what is done in Sec.~\ref{s.dispersionH}, one observes that the ${\bm k}$-mode profile on the ABC unit cell (see Fig.~\ref{Lieb_kagome}) has a form 
  \begin{equation}
  \bm{u}_{\bm{k}} = (0, a_{\bm{k}},-b_{\bm{k}}) \nonumber 
 \end{equation}
 while the $\bm{v}_{\bm{k}}$ coefficients  are identically equal to zero, where the numbering of the unit cell sites is as in Fig.~\ref{Lieb_kagome}. As expected from the mechanism of formation of the flat band (see Fig.~\ref{flatband-sketch}), its modes have only support on the link sites, with alternating signs which give rise to destructive interference in the propagation.  
 
  When studying the quench dynamics originated by initializing the system in the uniform MF ground state, one is faced with the interesting observation that the population of the flat band associated with such an initial state is identically \emph{zero} (see Table~\ref{betabosons_populations}), so that the flat band is completely projected out of the subsequent dynamics.
   Willing to explore the effects of the flat band on the dynamics, we must modify the quench protocol and focus on a global/local quench, in which the addition of a spin flip on top of the MF Ansatz can explicitly populate the flat band. We choose to initialize the dynamics from the state $b^{\dagger}_{l_0 p_0}|\Psi_{\rm MF}^{(-)}\rangle$  where $l_0 p_0$ are the coordinates of a link site, as link sites are the support of the localized modes belonging to the flat band.  
 The localized spin flip actually excites all three bands, because the modes on the dispersive bands have also a finite support on the link sites. Therefore, when monitoring the evolution of the spin-deviation profile, Eq.~\eqref{e.spindeviation}, one should expect that part of the initially localized magnetization propagates ballistically away from the site of origin, carried away by the modes on the dispersive bands. At the same time, a sizable fraction of the magnetic moment should remain bound to the neighborhood of the original site, as it is \emph{trapped} by the localized eigenmodes of the system which overlap with the site of interest. 
 
  This is indeed what is observed in Fig.~\ref{local_quench_Lieb}(a), showing clearly the caging effect of a fraction of the initially injected magnetization. The spatial profile of the localized fraction of the spin deviation clearly reconstructs the form of the localized modes \footnote{As we already pointed out in the introduction, the simple localized modes (or Aharonov-Bohm cages) shown in Fig.~\ref{flatband-sketch} are not orthogonal to each other. A proper basis of localized eigenmodes spanning the flat band is provided by a Wannier-function construction \cite{huber_bose_2010}, with resulting orthogonal localized modes which possess a thin tail beyond the localized structure of the Aharonov-Bohm cages.}, providing a precious real-space insight into the structure of the localized states, in a similar fashion to what has been observed recently in photonic crystals \cite{mukherjee_observation_2015,vicencio_observation_2015,mukherjeeetal2018}.  The study of this phenomenon using spin waves in quantum simulators of magnetism has the further advantage that the transverse field offers a continuous tuning knob for the localization effect.  
  On a more quantitative footing, Fig.~\ref{local_quench_Lieb}(b-d) shows the fraction of the initially injected quasiparticle that remains localized around the original site via the quantity (hereafter called \emph{localized fraction})
    \begin{equation}
    \mathcal{D}(t)=\sum_{i\in {\rm LM}(i_0)}  \delta m_{i}(t)
\end{equation}
where the sum runs over the sites of the two localized modes (LM) which overlap with the site  $i_0 = (l_0 p_0)$, as sketched in the inset of Fig.~\ref{local_quench_Lieb}(b). There we observe that the localized fraction of the magnetic moment follows a dynamics depending on the value of the transverse field, although it saturates to the same constant value at long times (Fig.~\ref{local_quench_Lieb}(c)). On the other hand, the transverse field controls the time at which the localized fraction saturates to its asymptotic value: Fig.~\ref{local_quench_Lieb}(d) shows that curves for different values of $\Gamma$ can be brought to collapse when time is expressed in units of $\bar{t}=d/(2v_{G,\mathrm{max}})$, where $v_{G,\rm max}$ is the maximum group velocity of all three bands.

\begin{center}
\begin{figure*}[t]
\includegraphics[width=0.7\textwidth]{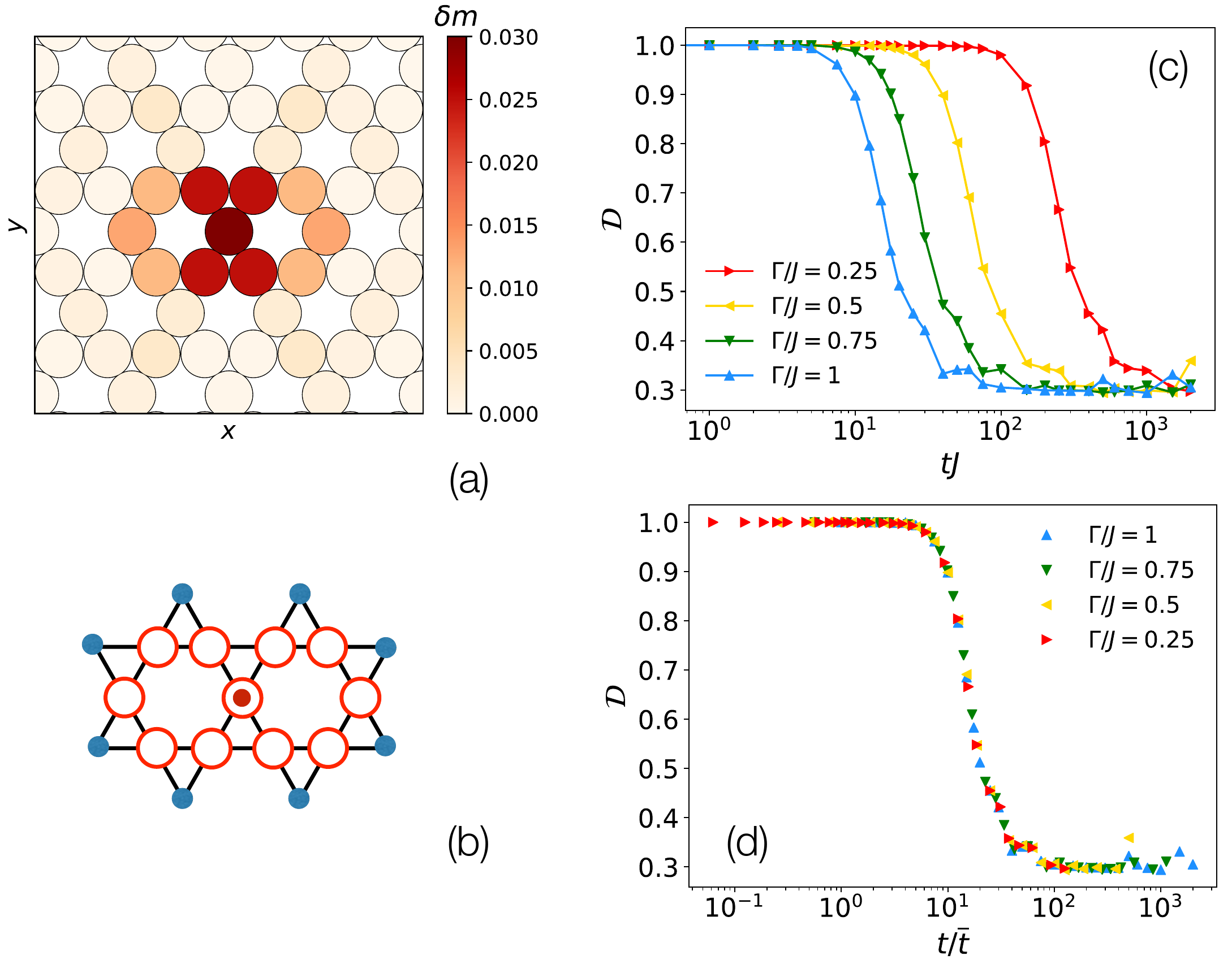}
\caption{Local/global quench on the Kagom\'e lattice. (a) Spin deviation profile $\delta m_{lp}$ at long times following a local/global quench. (b) Open dots indicate the region LM$(i_0)$ covered by the maximally localized modes which overlap with the site $i_0$ (indicated by a red dot).  (c) Evolution of the localized fraction of the spin deviation computed for a $30\times 30$ Kagom\'e lattice for different values of $\Gamma/J$. (d) Same quantity as in panel (c), plotted as a function of the characteristic time $\bar t$ (see main text).}
\label{local_quench_Kagome}
\end{figure*}
\end{center}

\subsection{Kagom\'e lattice: global and global/local quench}

At variance with the Lieb lattice, all sites of the Kagom\'e lattice are equivalent, so that MF ground state of $-{\cal H}$  is characterized by the same tilting angle $\theta = \arcsin (\Gamma/2J)$ everywhere. The LSW spectrum built around this state has the same main features as the spectrum of the tight-binding problem, namely two lower dispersive bands touching at a Dirac cone, and a perfectly flat upper band (which meets the middle band at a quadratic touching point, see Fig.~\ref{Lieb_kagome} (b)). 
The ${\bm k}$-modes associated with the flat band have the following structure on the unit cell (with sites numbered as in Fig.~\ref{Lieb_kagome}(b))
 \begin{equation}
  \bm{u}_{\bm{k}} = a_{\bm{k}}(1, 1,-1) \notag 
 \end{equation}
 and similarly for the $\bm{v}_{\bm{k}}$ coefficients ; the sign structure of the coefficients is responsible for the Aharonov-Bohm caging. 

A global quench in the Kagom\'e lattice, corresponding to the choice of $|\Psi_{\rm MF}\rangle$ as initial state, does not clearly show the presence of localized modes. Indeed, as one can notice in Table~\ref{betabosons_populations}, the order of magnitude between the populations of the lowest and flat bands are very different. The direct consequence is that the dynamics of correlations is largely dominated by the dispersive low-energy modes.

 On the other hand, a global/local quench starting from the state $b^{\dagger}_{l_0 p_0}|\Psi_{\rm MF}^{(-)}\rangle$ (where $l_0 p_0$ is an arbitrary site in the lattice) highlights the existence of  localized eigenmodes associated with the flat band -- see Fig.~\ref{local_quench_Kagome}. As seen in the case of the Lieb lattice, a fraction of the spin deviation initially imposed to the MF ground state remains trapped around the injection site, and it reconstructs the hexagonal shape of the two localized eigenmodes which overlap with such a site (Fig.~\ref{local_quench_Kagome}(a-b)).   
Quantifying again the fraction of the spin deviation which remains trapped on the two localized eigenmodes by the quantity ${\cal D}(t)$, we observe that, as function of the applied field, the localization fraction decreases toward a same limit value (Fig.~\ref{local_quench_Kagome}(c)).
All curves collapse onto one another after rescaling time by a characteristic time $\overline{t}=d/(2v_{G,\rm max})\propto\Gamma^{-2}$, where $v_{G,\rm max}$ is the maximum group velocity of all three bands (Fig.~\ref{local_quench_Kagome}(d)).

\begin{center}
\begin{figure}[t]
    \centering
    \label{f.vdW}
    \includegraphics[width=\columnwidth]{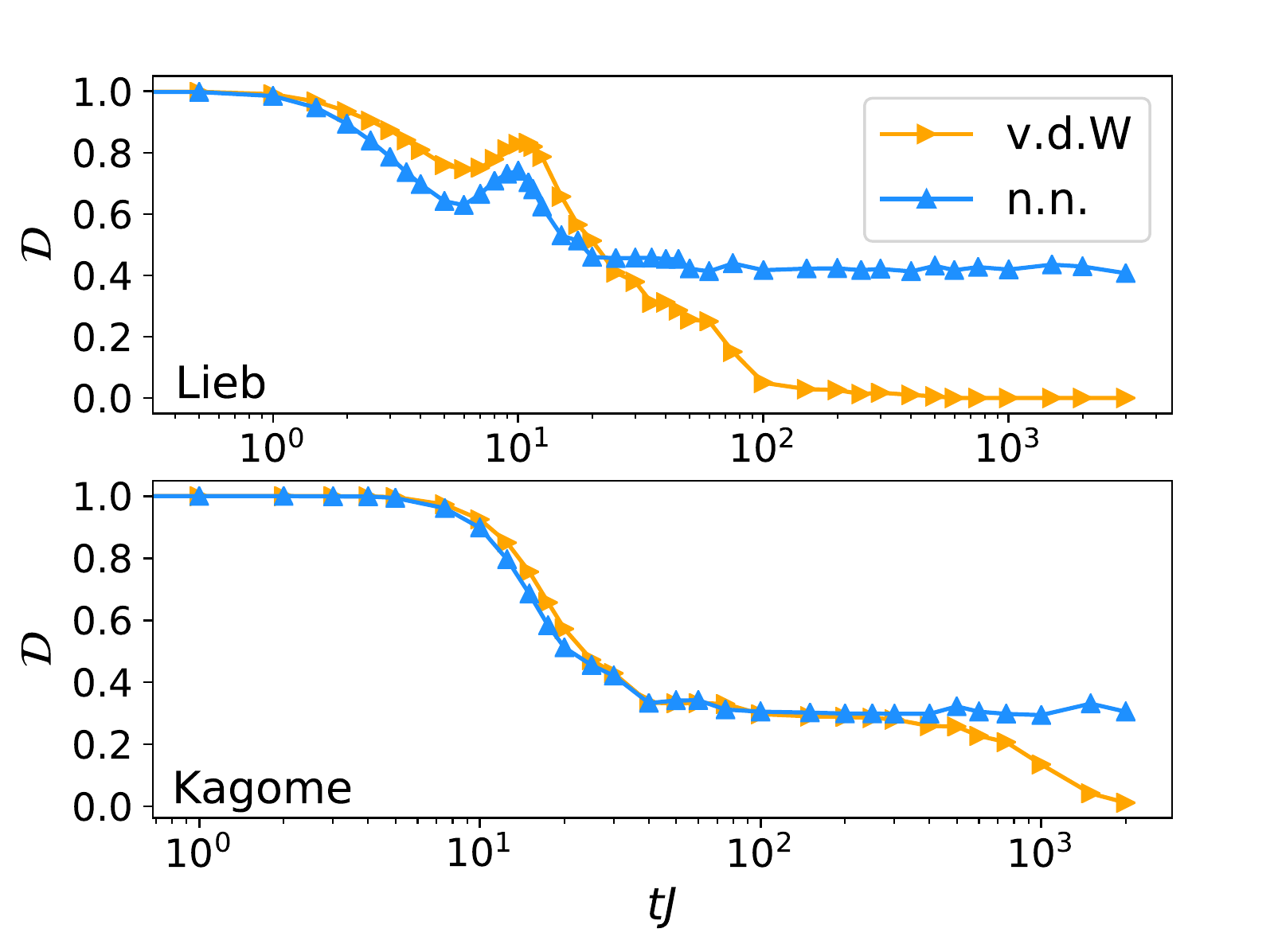}
    \caption{Time evolution of the localized fraction after a global/local quench in the presence of van-der-Waals (vdW) interactions, contrasted with the same quantity in the case of nearest-neighbour (nn) interaction. The calcultions are done on a Lieb lattice (a) and on a kagom\'e lattice (b). Both lattices have size $30\times 30$ with $\Gamma/J=-1$.}
\end{figure}
\end{center}

\subsection{Kagom\'e and Lieb lattice: var-der-Waals interactions}

The previous results are relevant to the physics of Rydberg atoms only under the condition that the above studied phenomena are robust to extending the nearest-neighbor interactions considered so far to the case of power-law, van-der-Waals interactions, decaying as $R^{-6}$ ($R$ being the inter-atomic distance). This condition can be explicitly testedx by computing the dynamics of the localized fraction in the presence of van-der-Waals interactions. The long-range nature of the intersite coupling inevitably introduces a finite bandwidth $\delta\omega$ to the flat band of the short-range-interacting model. As a consequence, the localized state acquires a finite lifetime $\tau \sim \delta\omega^{-1}$ beyond which it begins to delocalize. This effect is highlighted on Fig.~\ref{f.vdW} where, after a characteristic time $\tau$, the localized fraction computed with van-der-Waals interactions begins to depart from that calculated with nearest-neighbours interactions only, and to relax to zero. Interestingly, the "escape" time $\tau$ of the localized magnetic moment is much larger on the Kagom\'e lattice than on the Lieb lattice (all the other Hamiltonian parameters being the same), witnessing a smaller bandwidth induced by the power-law tail of the interactions. This can be naively understood by considering that, if $a$ is nearest-neighbor distance on both lattice, in the Lieb lattice the next-to-nearest neighbor is at distance $\sqrt{2} a$, while it is at distance $2a$ in the Kagom\'e lattice. Therefore the next-nearest-neighbor interaction term is 8 times larger on the Lieb lattice than on the Kagom\'e one. Nonetheless for both lattices the localized fraction remains very sizable up to times of the order $\sim 10 J^{-1}$, and specifically on the Kagom\'e lattice up to times $\gtrsim 5 *10^2 J^{-1}$. This opens a broad time window for an experimental observation of this Aharonov-Bohm caging effect under realistic conditions.

\subsection{Discussion}

 In this section we have shown that, due to the existence of flat bands in the Lieb and Kagom\'e lattice, the energy and magnetization which can be injected on a lattice site by a non-uniform quench (spin flip) can remain partially trapped around the site in question, and that this trapping phenomenon is rather robust to the presence of long-range interactions which emerge naturally in the case of Rydberg atoms. We remark that the fragmentation of the (integer) magnetic moment injected by a spin flip into a localized part and an extended part is a rather intriguing phenomenon of \emph{fractionalization} occurring at the level of linearized excitations. This is certainly not a fundamental fractionalization phenomenon, as the fractions into which the magnetic moment breaks up are not equal, namely they cannot constitute emergent elementary quasiparticles (to be contrasted with what is observed in one-dimensional systems \cite{Giamarchi-book} or in quantum spin liquids \cite{SavaryB2017}). The fragmentation phenomenon at hand can be viewed as a continuously tunable mechanism which entangle the region of injection of the magnetic moment with the rest of the system. An interesting question concerns the persistence of fragmentation of injected magnetic moments when going beyond LSW theory, and in particular when considering the injection of a macroscopic number of such localized magnetic moments in some arbitrary spatial pattern, leading to a macroscopic population in the flat band. While the stationary state of the evolution is most likely an ordinary thermalized state losing memory of the initial state, the magnetization pattern which is present in the initial state could persist over a significant time scale, thereby witnessing an important role of the flat bands even at the level of the full many-body dynamics. Exploring the possibility of such a phenomenon is obviously beyond the scopes of LSW theory, and we leave this investigation to future work.

\section{Conclusions and outlook}
\label{s.conclusion}

 Inspired by the potential of Rydberg quantum simulators of quantum magnetism, we have explored the quench dynamics of two-dimensional quantum Ising models possessing flat bands in the excitation spectrum. We have highlighted the impact of flat bands in the evolution of correlations and local observables after global and local quenches; and, in reverse, we have illustrated how quench dynamics can probe in a unique way the special nature of the states associated with the flat band. Indeed quench dynamics allows for a spectroscopic characterization of the flat-band modes in momentum/frequency space; and at the same time it provides a complementary real-space/real-time picture of the same modes, unveiling their spatial structure and their small or vanishing group velocity. This dual picture provides a unique way to characterize flat-band spectra, capitalizing on the full potential of Rydberg quantum simulators to implement quantum Ising models on arbitrary planar lattices, and to control and image such systems at the single-spin level.  
 
  In this work we have restricted our attention to the regime in which quantum spin dynamics can be linearized, and mapped onto that of a dilute gas of free bosonic quasiparticles -- this establishes an obvious link between the quantum simulation of flat-band systems in atomic physics, and the implementation of the same band structures using photons \cite{mukherjee_observation_2015,vicencio_observation_2015} or polaritons \cite{baboux_bosonic_2016}. Yet the use of $S=1/2$ spin systems offers the bonus that strong non-linearities are immediately achieved upon increasing the density of quasiparticles, as the latter are indeed hardcore repulsive. The peculiar spatial structure of the flat-band modes allows for inhomogeneous quench protocols that explicitly populate the flat band, potentially reaching the high-density regime. While accessing that regime is beyond the scopes of the present study, it certainly represents a very intriguing perspective. A fundamental question in that context concerns the stability of the quasiparticle population in the flat band against scattering processes into the other bands -- either conventional two-particle, three-particle etc. scattering, or conversion processes of \emph{e.g.} one quasiparticle in the flat band into two quasiparticles in dispersive bands and viceversa. The main tuning knobs of the model Hamiltonians of interest -- namely a longitudinal field (corresponding to the detuning) and a transverse field (corresponding to the Rabi frequency) coupling to the spins -- provide a way to alter the band structure while preserving the flat band: this may offer a way to control some of the decay mechanisms of the flat-band population. The non-linear dynamics of a macroscopic population of quasiparticles in the flat band raises fascinating questions -- such as that of the mechanism by which quasiparticles, initially prepared in a spatially inhomogeneous pattern, escape localization in the flat-band eigenmodes and relax towards a homogeneous thermal state.   
 
\section{Acknowledgments}
We thank A. Browaeys for insightful discussions.

\appendix
\section{Linear Spin-wave theory}
\label{app1}

The philosophy of spin-wave theory hinges on a treatment of the quantum fluctuations around the mean-field state of the Hamiltonian in terms of bosonic operators (following the prescription given in \ref{LSW}). The resulting Hamiltonian can be expanded up to an arbitrary order but a limitation to the second order enables an approximate description of the system as a collection of coupled harmonic oscillators. Thus, the Hamiltonian consists of two contributions, the first being the mean-field energy $E_{\mathrm{MF}}$ amounting to the energy of classical spins
\begin{align}
    E_{\mathrm{MF}}&=\dfrac{S^2}{2}\sum_{lp;l'p'}{J^{ll'}_{pp'}\cos\theta_{lp}\cos\theta_{l'p'}}\notag\\
    &-S\Gamma\sum_{lp}{\sin\theta_{lp}}-SH\sum_{lp}{\cos\theta_{lp}}.
\end{align}
The second contribution is a quadratic form of bosonic operators represented by the matrix elements of $\mathcal{A}_{lp;l'p'}$ as represented in Eq.\eqref{e.H2} with
\begin{equation}
(\mathcal{A}_{lp;l'p'})_{ij}=\dfrac{J^{ll'}_{pp'}}{2}-\dfrac{h_{lp}}{2}\delta_{ll'}\delta_{ij}
\end{equation}
and,
\begin{eqnarray}
    h_{lp}&=&S\sum_{l'p'} J^{ll'}_{pp'}\cos\theta_{lp}\cos\theta_{l'p'} \nonumber \\
    &~& -\Gamma\sin\theta_{lp} - H \cos\theta_{lp}.
\end{eqnarray}

Using the long-range order of the system, we perform a Fourier decomposition of the bosonic fields leading to a compact formulation of the quadratic part (Eq.\eqref{e.H2'}),
\begin{equation}
    \mathcal{H}_{\bm{k}}=\begin{pmatrix}
    A_{\bm{k}} & B_{\bm{k}}\\
    B_{\bm{k}} & A_{\bm{k}}\\
    \end{pmatrix}
\end{equation}
where $\left(B_{\bm{k}}\right)_{pp'}=s/2\sin\theta_p\sin\theta_{p'}\sum_l{e^{i\bm{k}\cdot\bm{r}_l}J_{pp'}^{0l}}$ and $\left(A_{\bm{k}}\right)_{pp'}=\left(B_{\bm{k}}\right)_{pp'}-h_{p}\delta_{pp'}$.

\section{Correlation function}
\label{app2}
Still using the linearized expressions of the spin-components, the correlation function defined in Eq.\eqref{e.corr} can be written in an approximate manner with the $b$ bosonic operators.
However, combined terms in $S^{z'}$ give rise to contributions of different orders in $b$-bosons. Using Wick's theorem, one can simplify and express the correlation functions only in terms of the two points correlators, such that
\begin{align}
\label{corr_app}
    C^{xx}(\bm{r}_{lp},\bm{r}_{l'p'};t)&=\sin\theta_{p}\sin\theta_{p'}\left(\langle b^{\dagger}_{lp}b_{l'p'}\rangle\langle b_{lp}b^{\dagger}_{l'p'}\rangle\right.\notag\\
    &+\left.\langle b_{lp}b_{l'p'}\rangle\langle b^{\dagger}_{lp}b^{\dagger}_{l'p'}\rangle\right)\notag\\
    &+\dfrac{s}{2}\cos\theta_p\cos\theta_{p'}\left(\langle b^{\dagger}_{lp}b_{l'p'}\rangle+\langle b_{lp}b^{\dagger}_{l'p'}\rangle\right.\notag\\
    &+\left.\langle b_{lp}b_{l'p'}\rangle+\langle b^{\dagger}_{lp}b^{\dagger}_{l'p'}\rangle\right).
\end{align}

\section{Quench spectroscopy}

In this section we focus on the time evolution of the momentum-dependent structure factor, providing the derivation of the quench-spectroscopy formula, Eq.~\eqref{corr_kspace}. Following the convention taken in Eq.\eqref{fourier_convention}, the $\bm{k}$-space correlation function is expressed as
\begin{align}
    S^{xx}_{pp'}\left(\bm{k},t\right)&\simeq\sin\theta_{p}\sin\theta_{p'}\sum_{\bm{q}}\left(\langle b^\dagger_{-\bm{k}+\bm{q},p}b_{-\bm{k}+\bm{q},p'}\rangle\langle b_{\bm{q},p}b^\dagger_{\bm{q},p'}\rangle\right.\notag\\
    &+\left.\langle b^\dagger_{-\bm{k}+\bm{q},p}b^\dagger_{\bm{k}-\bm{q},p'}\rangle\langle b_{\bm{q},p}b_{-\bm{q},p'}\rangle\right)\notag\\
    &+\dfrac{s}{2}\cos\theta_p\cos\theta_{p'}\left(\langle b^\dagger_{-\bm{k},p}b_{-\bm{k},p'}\rangle+\langle b_{\bm{k},p}b^\dagger_{\bm{k},p'}\rangle\right.\notag\\
    &+\left.\langle b^\dagger_{-\bm{k},p}b^\dagger_{\bm{k},p'}\rangle+\langle b_{\bm{k},p}b_{-\bm{k},p'}\rangle\right).
\end{align}
Given the diluteness of the quasi-particle gas, we can safely neglect the quartic terms, and restrict our attention to the quadratic terms only. Notice that the numerical calculation leading
to Fig.~\ref{quench_spectroscopy} includes the quartic terms as well; the success of our analysis based uniquely on the quadratic terms confirms the weakness of the quartic terms.  

 The time Fourier transform of the quadratic terms  can be explicitly computed and, it carries the spectral information we are searching for.
Indeed, the Bogoliubov transformation leads to an expression of the bosonic excitation operator $b_{\bm{k},p}$ in terms of oscillating functions,
\begin{align}
    b_{\bm{k},p}(t)&=\sum_{r,s}\left[(u^{(r) *}_{\bm{k},p}u^{(r)}_{\bm{k},s}-v^{(r)}_{\bm{k},p}v^{(r)*}_{\bm{k},s})\cos(\omega^{(r)}_{\bm{k}}t)\right.\notag\\
    &\left.-i(u^{(r) *}_{\bm{k},p}u^{(r)}_{\bm{k},s}+v^{(r)}_{\bm{k},p}v^{(r)*}_{\bm{k},s})\sin(\omega^{(r)}_{\bm{k}}t)\right]b_{\bm{k},s}(0)\notag\\
    &+\left[(u^{(r) *}_{\bm{k},p}v^{(r)}_{\bm{k},s}-v^{(r)}_{\bm{k},p}u^{(r)*}_{\bm{k},s})\cos(\omega^{(r)}_{\bm{k}}t)\right.\notag\\
    &-i\left.(u^{(r) *}_{\bm{k},p}v^{(r)}_{\bm{k},s}+v^{(r)}_{\bm{k},p}u^{(r)*}_{\bm{k},s})\sin(\omega^{(r)}_{\bm{k}}t)\right]b^\dagger_{-\bm{k},s}(0).
\end{align}

As a result, the time dependence of the structure factor is provided by a sum of oscillating functions, as summarized in Eq.~\eqref{corr_kspace}, with the following prefactors:

\begin{align}
 f_{\bm{k},pp'}&=\dfrac{s}{2}\cos\theta_p\cos\theta_{p'}\left( \delta_{pp'}\right.\notag\\
 &+\sum_{r}\left[2\left(u^{(r)*}_{\bm{k},p}u^{(r)}_{\bm{k},p'}\vert\bm{v}^{(r)}_{\bm{k}}\vert^2+v^{(r)}_{\bm{k},p}v^{(r)*}_{\bm{k},p'}\vert\bm{u}^{(r)}_{\bm{k}}\vert^2\right)\right.\notag\\
 &-\left(v^{(r)}_{\bm{k},p}u^{(r)}_{\bm{k},p'}\left((\bm{u}^{r}_{\bm{k}}\cdot \bm{u}^{r}_{\bm{k}})+(\bm{v}^{r}_{\bm{k}}\cdot \bm{v}^{r}_{\bm{k}})\right)^*\right.\notag\\
  &+\left.\left.\left.u^{(r)*}_{\bm{k},p}v^{(r)*}_{\bm{k},p'}\left((\bm{u}^{r}_{\bm{k}}\cdot \bm{u}^{r}_{\bm{k}})+(\bm{v}^{r}_{\bm{k}}\cdot \bm{v}^{r}_{\bm{k}})\right)\right)\right]\right),
\end{align}

\begin{align}
 g^{rr'}_{pp'}(\bm{k})&=\dfrac{s}{2}\cos\theta_p\cos\theta_{p'}\left[-2\left( v^{(r)}_{\bm{k},p}u^{(r')}_{\bm{k},p'}(\bm{u}^{(r)}_{\bm{k}}\cdot\bm{v}^{(r')}_{\bm{k}})^*\right.\right.\notag\\
 &+v^{(r')}_{\bm{k},p}u^{(r)}_{\bm{k},p'}(\bm{u}^{(r')}_{\bm{k}}\cdot\bm{v}^{(r)}_{\bm{k}})^*\notag\\
 &+\left. u^{(r)*}_{\bm{k},p}v^{(r')*}_{\bm{k},p'}(\bm{v}^{(r)}_{\bm{k}}\cdot\bm{u}^{(r')}_{\bm{k}})+u^{(r')*}_{\bm{k},p}v^{(r)*}_{\bm{k},p'}(\bm{v}^{(r')}_{\bm{k}}\cdot\bm{u}^{(r)}_{\bm{k}}) \right)\notag\\
 &+\left((\bm{u}^{(r)}_{\bm{k}}\cdot\bm{v}^{(r)*}_{\bm{k}}+\bm{v}^{(r)}_{\bm{k}}\cdot\bm{u}^{(r)*}_{\bm{k}})(u^{(r)*}_{\bm{k},p}u^{(r')}_{\bm{k},p'}+v^{(r')}_{\bm{k},p}v^{(r)*}_{\bm{k},p'})\right.\notag\\
 &+\left.\left. (\bm{v}^{(r)*}_{\bm{k}} \cdot \bm{u}^{(r')}_{\bm{k}} + \bm{v}^{(r')}_{\bm{k}} \cdot \bm{u}^{(r)*}_{\bm{k}}) (u^{(r')*}_{\bm{k},p} u^{(r)}_{\bm{k},p'} + v^{(r)}_{\bm{k},p} v^{(r')*}_{\bm{k},p'})\right)\right],
\end{align}

\begin{align}
 h^{rr'}_{pp'}(\bm{k})&=\dfrac{s}{2}\cos\theta_p\cos\theta_{p'}\left[ 2\left( u^{(r)*}_{\bm{k},p} u^{(r')}_{\bm{k},p'} (\bm{v}^{(r)}_{\bm{k}} \cdot \bm{v}^{(r')*}_{\bm{k}})\right.\right.\notag\\
 &+ u^{(r')*}_{\bm{k},p} u^{(r)}_{\bm{k},p'} (\bm{v}^{(r')}_{\bm{k}} \cdot \bm{v}^{(r)*}_{\bm{k}}) \notag\\
 & +\left. v^{(r)}_{\bm{k},p}v^{(r')*}_{\bm{k},p'} (\bm{u}^{(r)*}_{\bm{k}} \cdot \bm{u}^{(r')}_{\bm{k}}) + v^{(r')}_{\bm{k},p} v^{(r)*}_{\bm{k},p'} (\bm{u}^{(r')*}_{\bm{k}}\cdot\bm{u}^{(r)}_{\bm{k}}) \right)\notag\\
 &-\left((\bm{u}^{(r)}_{\bm{k}}\cdot\bm{u}^{(r')}_{\bm{k}}+\bm{v}^{(r)}_{\bm{k}}\cdot\bm{v}^{(r)}_{\bm{k}})^*(v^{(r)}_{\bm{k},p}u^{(r')}_{\bm{k},p'}+v^{(r')}_{\bm{k},p}u^{(r)}_{\bm{k},p'})\right.\notag\\
 &+\left.\left. (\bm{u}^{(r)}_{\bm{k}} \cdot \bm{u}^{(r')}_{\bm{k}} + \bm{v}^{(r)}_{\bm{k}} \cdot \bm{v}^{(r')}_{\bm{k}}) (u^{(r)}_{\bm{k},p} v^{(r')}_{\bm{k},p'} + u^{(r')}_{\bm{k},p} v^{(r)}_{\bm{k},p'})\right)^*\right],
\end{align}

\begin{align}
 \overline{g}^{rr'}_{pp'}(\bm{k})&=\dfrac{s}{2}\cos\theta_p\cos\theta_{p'}\left[2\left( v^{(r)}_{\bm{k},p} u^{(r')}_{\bm{k},p'} (\bm{u}^{(r)}_{\bm{k}} \cdot \bm{v}^{(r')}_{\bm{k}})^* \right.\right.\notag\\
 &- u^{(r)*}_{\bm{k},p} v^{(r')*}_{\bm{k},p'} (\bm{v}^{(r)}_{\bm{k}} \cdot \bm{u}^{(r')}_{\bm{k}})\notag\\
 & +\left. v^{(r')}_{\bm{k},p} u^{(r)}_{\bm{k},p'} (\bm{v}^{(r)}_{\bm{k}} \cdot \bm{u}^{(r')}_{\bm{k}})^* - u^{(r')*}_{\bm{k},p} v^{(r)*}_{\bm{k},p'} (\bm{v}^{(r')}_{\bm{k}} \cdot \bm{u}^{(r)}_{\bm{k}}) \right)\notag\\
 &-\left((\bm{u}^{(r)}_{\bm{k}}\cdot\bm{v}^{(r')*}_{\bm{k}})(u^{(r)*}_{\bm{k},p} u^{(r')}_{\bm{k},p'}-v^{(r')}_{\bm{k},p} v^{(r)*}_{\bm{k},p'})\right.\notag\\
 &+\left.(\bm{u}^{(r')}_{\bm{k}}\cdot\bm{v}^{(r)*}_{\bm{k}})(u^{(r')*}_{\bm{k},p} u^{(r)}_{\bm{k},p'}-v^{(r)}_{\bm{k},p} v^{(r')*}_{\bm{k},p'})\right)\notag\\
 &+\left((\bm{u}^{(r)*}_{\bm{k}}\cdot\bm{v}^{(r')}_{\bm{k}})(u^{(r')*}_{\bm{k},p} u^{(r)}_{\bm{k},p'}-v^{(r)}_{\bm{k},p} v^{(r')*}_{\bm{k},p'})\right.\notag\\
 &+\left.\left.(\bm{u}^{(r')*}_{\bm{k}}\cdot\bm{v}^{(r)}_{\bm{k}})(u^{(r)*}_{\bm{k},p} u^{(r')}_{\bm{k},p'}-v^{(r')}_{\bm{k},p} v^{(r)*}_{\bm{k},p'})\right)\right],
\end{align}

and finally

\begin{align}
 \overline{h}^{rr'}_{pp'}(\bm{k})&=\dfrac{s}{2}\cos\theta_p\cos\theta_{p'}\left[ 2\left( u^{(r)*}_{\bm{k},p} u^{(r')}_{\bm{k},p'} (\bm{v}^{(r)}_{\bm{k}} \cdot \bm{v}^{(r')*}_{\bm{k}})\right.\right.\notag\\
 &- v^{(r)}_{\bm{k},p} v^{(r')*}_{\bm{k},p'} (\bm{u}^{(r)*}_{\bm{k}} \cdot \bm{u}^{(r')}_{\bm{k}})\notag\\
 & -\left. u^{(r')*}_{\bm{k},p} u^{(r)}_{\bm{k},p'} (\bm{v}^{(r')}_{\bm{k}} \cdot \bm{v}^{(r)*}_{\bm{k}})+ v^{(r')}_{\bm{k},p} v^{(r)*}_{\bm{k},p'} (\bm{u}^{(r')*}_{\bm{k}} \cdot \bm{u}^{(r)}_{\bm{k}}) \right)\notag\\
 &-\left((\bm{v}^{(r)}_{\bm{k}}\cdot\bm{v}^{(r')}_{\bm{k}})^* (v^{(r)}_{\bm{k},p} u^{(r')}_{\bm{k},p'}-v^{(r')}_{\bm{k},p} u^{(r)}_{\bm{k},p'})\right.\notag\\
 &-\left.(\bm{u}^{(r')}_{\bm{k}}\cdot\bm{u}^{(r)}_{\bm{k}})(u^{(r)*}_{\bm{k},p} v^{(r')*}_{\bm{k},p'}-u^{(r')*}_{\bm{k},p} v^{(r)*}_{\bm{k},p'})\right)\notag\\
 &+\left((\bm{u}^{(r)}_{\bm{k}}\cdot\bm{u}^{(r')}_{\bm{k}})^*(u^{(r)}_{\bm{k},p} v^{(r')}_{\bm{k},p'}-u^{(r')}_{\bm{k},p} v^{(r)}_{\bm{k},p'})\right.\notag\\
 &-\left.\left.(\bm{v}^{(r)}_{\bm{k}}\cdot\bm{v}^{(r')}_{\bm{k}})(v^{(r)}_{\bm{k},p} u^{(r')}_{\bm{k},p'}-v^{(r')}_{\bm{k},p} u^{(r)}_{\bm{k},p'})^*\right)\right].
\end{align}

Notice that the $\bar{g}$ and $\bar{h}$ coefficients are purely imaginary, resulting in a real expression in Eq.~\eqref{corr_kspace}. 

\bibliography{biblio_flatband}

\end{document}